\documentclass[12pt]{article}
\usepackage{epsf}
\def\beq{\begin{equation}}
\def\eeq{\end{equation}}
\def\bea{\begin{eqnarray}}
\def\eea{\end{eqnarray}}

\def\nnb{\nonumber}

\def\aga{\left\{}
\def\adr{\right\}}

\def\nnb{\nonumber}

\def\ba{\begin{array}}
\def\ea{\end{array}}
\def\bea{\begin{eqnarray}}
\def\eea{\end{eqnarray}}

\title{ {\bf
$b\rightarrow s \tau^+ \tau^-$  decay in the two Higgs doublet model with flavor 
changing neutral currents}}
\author{\vspace{1cm}\\
         {\bf E. O. Iltan} 
         \thanks{E-mail address:
        eiltan@heraklit.physics.metu.edu.tr}
 \\
         Physics Department, Middle East Technical University \\
         Ankara, Turkey\\        \vspace{5mm}\\
        {\bf G. Turan}
        \thanks{E-mail address:
        gsevgur@rorqual.metu.edu.tr}
 \\
        Physics Department, Middle East Technical University \\
        Ankara, Turkey\\}

\date{}

\begin{document}
\setlength{\baselineskip}{24pt}
\maketitle
\setlength{\baselineskip}{7mm}
\begin{abstract}
We study the decay width and forward-backward asymmetry of the lepton pair   
for the inclusive decay $b\rightarrow s \tau^+ \tau^-$ in the two 
Higgs doublet model with three level flavor changing neutral currents 
(model III) and analyse the dependencies of these quantities on the  
model III  parameters, including the leading order QCD corrections. 
We found that there is a considerable enhancement in the decay width and 
neutral Higgs effects are detectable for large values of the parameter 
$\bar{\xi}_{N,\tau\tau}^D$.
\end{abstract} 
\thispagestyle{empty}
\newpage
\setcounter{page}{1}
\section{Introduction}
Currently, there is an impressive experimental effort for studying rare
B-meson decays at  SLAC (BaBar), KEK (BELLE), B-Factories, DESY (HERA-B)
since these decays are rich phenomenologically.  They are induced 
by flavor changing neutral currents (FCNC) at loop level in the Standard
model (SM) and with 
the forthcoming experiments, it would be possible to test the flavour sector 
of the SM in a high precision, as well as to reveal the physics beyond, such 
as two Higgs Doublet model (2HDM), Minimal Supersymmetric extension of 
the SM (MSSM) \cite{Hewett}, etc. 

Among the rare B decays, $B\rightarrow K^* l^+ l^-$ process has received a 
great interest since the SM prediction for its branching ratio ($Br$) is
large enough to be  measured in the near future. 
This decay is induced by $b\rightarrow s l^+l^-$ transition 
at the quark level and in the literature it 
has been investigated extensively for $l=e,\mu$  in  the SM, 2HDM and MSSM 
\cite{R4}- \cite{R10}.
When $l=e,\mu$, the neutral Higgs boson (NHB) effects are safely neglected
in the 2HDM because they enter in the expressions with the factor 
$m_{e(\mu)}/m_W$. However, for $l=\tau$, this factor is not negligible and 
NHB effects can give important contribution. In \cite{Chao,Logan}, 
$B\rightarrow X_s\tau^+\tau^-$ process was studied in the 2HDM and it was
shown that NHB effects are sizable  for large values of 
$tan\beta$.

In this work, we study the $b\rightarrow s\tau^+\tau^-$ decay 
in the general 2HDM , so-called model III. We include NHB effects and 
make the full calculation using the on-shell renormalization prescription. 
We investigate the dependencies of  the differential decay width 
$d\Gamma/ds$ and the  decay width 
$\Gamma$ on the scale invariant lepton mass square "$s$" and some model III 
parameters, namely $m_{H^{\pm}}$, $\bar{\xi}_{N,bb}^D$ and  
$\bar{\xi}_{N,\tau\tau}^D$.
Further, we calculate the differential (direct) forward-backward asymmetry 
$A_{FB}(s)$ ($A_{FB}$) of the lepton pair in terms of
the above parameters. We show that a large enhancement 
is possible in the decay width of the process $b\rightarrow s\tau^+\tau^-$
for some values of the model III parameters and NHB
effects become considerable  for large values of $\bar{\xi}_{N,\tau\tau}^D$.   

The paper is organized as follows:
In Section 2, we present the leading order (LO) QCD corrected effective
Hamiltonian  and the corresponding matrix element for the inclusive 
$b\rightarrow s\tau^+\tau^-$ decay. Further, we give the expression for 
$A_{FB}(s)$ and $A_{FB}$ of the lepton pair. Section 3 is devoted to the 
analysis of the  new Wilson coefficients coming from the NHB effects and 
the  dependencies of $d\Gamma/ds$, $\Gamma$, $A_{FB}(s)$ and $A_{FB}$ on 
the the Yukawa couplings $\bar{\xi}_{N,bb}^{D}$, $\bar{\xi}_{N,\tau\tau}^{D}$, 
the charged Higgs 
mass $m_{H^{\pm}}$, the parameter s and to the discussion of our results. 
In Appendices, we give the explicit forms of the operators appearing in the 
effective Hamiltonian and the corresponding Wilson coefficients.
\section{\bf The inclusive $b\rightarrow s \tau^+ \tau^-$ decay in the model 
III }
Model III (2HDM) permits the flavour changing neutral currents in the tree
level and the prize is various new parameters, i.e. Yukawa couplings. 
These couplings are responsible for the interaction of
quarks and leptons with gauge bosons, namely, the Yukawa interaction 
and in this general case it  reads as 
\begin{eqnarray}
{\cal{L}}_{Y}=\eta^{U}_{ij} \bar{Q}_{i L} \tilde{\phi_{1}} U_{j R}+
\eta^{D}_{ij} \bar{Q}_{i L} \phi_{1} D_{j R}+
\xi^{U}_{ij} \bar{Q}_{i L} \tilde{\phi_{2}} U_{j R}+
\xi^{D}_{ij} \bar{Q}_{i L} \phi_{2} D_{j R} + h.c. \,\,\, ,
\label{lagrangian}
\end{eqnarray}
where $L$ and $R$ denote chiral projections $L(R)=1/2(1\mp \gamma_5)$,
$\phi_{k}$, for $k=1,2$, are the two scalar doublets, $Q_{i L}$ are
quark and lepton doublets, $U_{j R}$, $D_{j R}$ are the corresponding 
singlets, $\eta^{U,D}_{ij}$, and $\xi^{U,D}_{ij}$ are the matrices of the 
Yukawa couplings. The Flavor changing (FC) part of the interaction is given 
by
\begin{eqnarray}
{\cal{L}}_{Y,FC}=
\xi^{U}_{ij} \bar{Q}_{i L} \tilde{\phi_{2}} U_{j R}+
\xi^{D}_{ij} \bar{Q}_{i L} \phi_{2} D_{j R} + h.c. \,\, .
\label{lagrangianFC}
\end{eqnarray}
The choice of $\phi_1$ and $\phi_2$
\begin{eqnarray}
\phi_{1}=\frac{1}{\sqrt{2}}\left[\left(\begin{array}{c c} 
0\\v+H^{0}\end{array}\right)\; + \left(\begin{array}{c c} 
\sqrt{2} \chi^{+}\\ i \chi^{0}\end{array}\right) \right]\, ; 
\phi_{2}=\frac{1}{\sqrt{2}}\left(\begin{array}{c c} 
\sqrt{2} H^{+}\\ H_1+i H_2 \end{array}\right) \,\, .
\label{choice}
\end{eqnarray}
with the vacuum expectation values,  
\begin{eqnarray}
<\phi_{1}>=\frac{1}{\sqrt{2}}\left(\begin{array}{c c} 
0\\v\end{array}\right) \,  \, ; 
<\phi_{2}>=0 \,\, ,
\label{choice2}
\end{eqnarray}
ensures decoupling of the SM and beyond.
In eq.(\ref{lagrangianFC}) the couplings  $\xi^{U,D}$ for the FC charged 
interactions are 
\begin{eqnarray}
\xi^{U}_{ch}&=& \xi_{neutral} \,\, V_{CKM} \nonumber \,\, ,\\
\xi^{D}_{ch}&=& V_{CKM} \,\, \xi_{neutral} \,\, ,
\label{ksi1} 
\end{eqnarray}
where  $\xi^{U,D}_{neutral}$ 
\footnote{In all next discussion we denote $\xi^{U,D}_{neutral}$ 
as $\xi^{U,D}_{N}$.} 
is defined by the expression
\begin{eqnarray}
\xi^{U,D}_{N}=(V_L^{U,D})^{-1} \xi^{U,D} V_R^{U,D}\,\, .
\label{ksineut}
\end{eqnarray}
Here the charged couplings appear as linear combinations of neutral 
couplings multiplied by $V_{CKM}$ matrix elements (see \cite{atwood} for
details). 

Now we would like to start with the calculation of the matrix element for 
the inclusive $b\rightarrow s \tau^+ \tau^-$  decay. The procedure is to 
integrate out the heavy degrees of freedom, namely $t$ quark, $W^{\pm}, 
H^{\pm}, H^0, H_{1}$, and $H_{2}$ bosons in the present case and obtain the 
effective theory. Here $H^{\pm}$ denote charged, $H^0$, $H_{1}$ and $H_{2}$ 
denote neutral Higgs bosons. Note that $H_{1}$ and $H_{2}$ are the same as 
the mass eigenstates $h^{0}$ and $A^{0}$ in the model III respectively, due 
to the choice given by eq. (\ref{choice}). The QCD corrections are done 
through matching the full  theory with the effective low energy one at 
the high scale $\mu=m_{W}$ and evaluating the Wilson coefficients from 
$m_{W}$ down to the lower scale $\mu\sim O(m_{b})$. In the model III 
(similar to the models I and II, 2HDM) neutral Higgs  particles bring new 
contributions to the matrix element of the process 
$b\rightarrow s \tau^+ \tau^-$ (see eq.(\ref{NHB})) since they enter in the 
expressions with the mass of $\tau$ lepton or related Yukawa coupling 
$\bar{\xi}^{D}_{N,\tau\tau}$. As being different from the model I and II, 
in the model III, there exist additional operators which are the flipped 
chirality partners of the former ones. However, the effects of the latter 
are negligible since the corresponding Wilson coefficients are small  due 
to the discussion given in section 3. Therefore, the effective Hamiltonian 
relevant for the process $b\rightarrow s \tau^+\tau^-$ is
\begin{eqnarray}
{\cal{H}}_{eff}=-4 \frac{G_{F}}{\sqrt{2}} V_{tb} V^{*}_{ts}\aga 
\sum_{i}C_{i}(\mu) O_{i}(\mu)+\sum_{i}C_{Q_i}(\mu) Q_{i}(\mu)\adr 
\, \, ,
\label{hamilton}
\end{eqnarray}
where $O_{i}$ are current-current ($i=1,2$), penguin ($i=3,...,6$),
magnetic penguin ($i=7,8$) and semileptonic ($i=9,10$) operators. 
Here, $C_{i}(\mu)$ are Wilson coefficients normalized at the
scale $\mu $ and given  in Appendix B. The additional operators $Q_{i}
(i=1,..,10)$ are due to the NHB exchange diagrams and $C_{Q_i}(\mu)$ are
their Wilson coefficients (see Appendices A and B) . 

During the calculations of NHB
contributions, we use the on-shell renormalization scheme
to overcome the logarithmic divergences. Taking the vertex function 
\begin{eqnarray}
\Gamma_{neutr}^{Ren}(p^2)=\Gamma_{neutr}^{0}(p^2)+\Gamma_{neutr}^{C} ,  
\label{vertren}
\end{eqnarray}
and using the renormalization condition 
\begin{eqnarray}
\Gamma_{neutr}^{Ren}(p^2=m^2_{neutr})=0 ,
\label{rencond}
\end{eqnarray}
we get the counter terms and then  calculate $\Gamma_{neutr}^{Ren}(p^2)$ . 
Here the phrase $neutr$ denotes the neutral Higgs 
bosons, $H^0$, $h^0$ and  $A^0$ and  $p$ is the momentum transfer.
 
Now we give the QCD corrected amplitude for the inclusive 
$b\rightarrow s \tau^+ \tau^-$ decay in the model III, 
\bea
{\cal M} & = & \frac{\alpha G_F}{ \sqrt{2}\, \pi} V_{tb} V_{ts}^*\Bigg{\{} 
C_9^{eff} (\bar s \gamma_\mu P_L b) \, \bar \tau \gamma^\mu \tau +
C_{10} ( \bar s \gamma_\mu P_L b) \, \bar \tau \gamma^\mu \gamma_5 \tau  \nnb 
\\ & -& 2 C_7 \frac{m_b}{p^2} (\bar s i \sigma_{\mu \nu} p^\nu P_R b) 
\bar \tau \gamma^\mu \tau + C_{Q_{1}}(\bar s  P_R b) \bar \tau  
\tau +C_{Q_{2}} (\bar s  P_R b) \bar \tau \gamma_5 \tau \Bigg{\}}~. 
\label{M1}
\eea
Using Eq.(\ref{M1}), the differential decay rate reads as
\bea
\frac{d \Gamma (b\rightarrow s \tau^{+}\tau^{-})}{d s} & = & 
Br(B\rightarrow X_{c}\ell \bar\nu) \frac{\alpha^2}{4\pi^2 f(m_{c}/m_{b})}
(1-s)^2 \left(1-\frac{4 t^2}{s} \right)^{1/2} 
\frac{|V_{tb} V_{ts}^*|^2}{|V_{cb}|^2} D(s) ,
\eea
with
\bea
D(s) & = &  |C^{eff}_{9}|^2 \left(1+\frac{2 t^2}{s} \right)
(1+2 s)+
4 |C_{7}|^{2} \left( 1+\frac{2 t^2}{s} \right) 
\left(1+\frac{2}{s} \right)+
|C_{10}|^2 \left[ 1+2 s+ \frac{2 t^2}{s} (1-4 s) \right]  \nnb \\
& + &  12 Re(C_{7} C^{eff *}_{9}) \left(1+\frac{2 t^2}{s} \right)^{1/2}+ 
\frac{3}{2} |C_{Q_{1}}|^2 (s-4 t^2)+ \frac{3}{2} |C_{Q_{2}}|^2 +
6 Re(C_{10} C^{*}_{Q_{2}}) ,
\eea
where $s=p^2/m^{2}_{b}$, $t=m_{\tau}/m_{b}$, and $f(x)$ is the phase-space 
factor given 
by $f(x)=1-8 x^2+8 x^6-x^8-24 x^4 \log x$. In the above expression for the
differential decay rate, we use the inclusive one since,
in the heavy quark effective theory,
the leading terms of inclusive decay rates of the heavy
hadrons  in the $1/m_b$ expansion becomes that of the free 
heavy quark, $b$-quark in our context.  

The forward-backward asymmetry  $A_{FB}$ of the lepton pair  is another
physical quantity which can be observed in the experiments and provide
important clues to test the theoretical models used. Using the definition 
of differential  $A_{FB}$  
\bea
A_{FB}(s) & = & \frac{ \int^{1}_{0}dz \frac{d \Gamma }{ds dz} - 
\int^{0}_{-1}dz \frac{d \Gamma }{ds dz}}{\int^{1}_{0}dz 
\frac{d \Gamma }{ds dz}+ \int^{0}_{-1}dz \frac{d \Gamma }{ds dz}}
\label{AFB1} 
\eea
with $z=\cos \theta$, where   $\theta$ is the angle between the momentum of 
the b-quark and that of $\tau^{+}$ in the center of mass frame of the 
dileptons $\tau^{+}\tau^{-}$, we get
\bea
A_{FB}(s)=\frac{E(s)}{D(s)}. 
\label{AFB2}
\eea
Here,
\bea
E(s) & = & Re(C^{eff}_{9}C^*_{10} s+ 2 C_{7}C^*_{10}+ 
C^{eff}_{9}C^{*}_{Q_{1}}t+2 C_{7}C^{*}_{Q_2}t)
\eea
In addition,  $A_{FB}$ can be defined as
\bea
A_{FB} & = & \frac{\int^{1}_{0}dz \frac{d \Gamma }{dz} - 
\int^{0}_{-1}dz \frac{d \Gamma }{dz}}{\Gamma }.\label{AFB3} 
\eea

Note that during the calculations of $\Gamma $ and $A_{FB}$, we take into
account only the second resonance for the LD effects coming from the
reaction $b \rightarrow s \psi_i \rightarrow s \tau^{+}\tau^{-}$, where 
$i=1,..,6$ and divide the
integration region for $s$ into two parts : $\frac{4 m^2_{\tau}}{m^2_b}\leq
s \leq \frac{(m_{\psi_2}-0.02)^2}{m^2_b}$ and 
$\frac{(m_{\psi_2}+0.02)^2}{m^2_b}\leq s \leq 1$, where
$m_{\psi_2}=3.686\,GeV$ is the mass of the second resonance 
(see Appendix B for LD contributions). 

\section{Discussion}
In the general 2HDM model, there are many free parameters, such as 
masses of charged and neutral Higgs bosons and the complex Yukawa couplings, 
$\xi_{ij}^{U,D}$, where $i,j$ are quark flavor indices and these parameters 
should be restricted using the experimental measurements. Usually, the 
stronger restrictions to the new couplings are obtained from the analysis 
of the $\Delta F=2$ (here $F=K,B_d,D$) decays, the $\rho$ parameter and the 
$B\rightarrow X_s \gamma$ decay. 

The neutral Higgs bosons $h_0$ and $A_0$ give contributions to the Wilson 
coefficient $C_7$ (see the appendix of \cite{alil2} for details)   
\begin{eqnarray}
C_7^{h_0}(m_W)&=& (V_{tb} V^{*}_{ts} )^{-1}\sum_{i=d,s,b} \bar{\xi}^{D}_{N,bi} 
\,\,\bar{\xi}^{D}_{N,is}\,  \frac{Q_i}{8\, m_i\, m_b}
\nonumber \,\,, \\ 
C_7^{A_0}(m_W)&=& (V_{tb} V^{*}_{ts} )^{-1}\sum_{i=d,s,b} \bar{\xi}^{D}_{N,bi} 
\,\, \bar{\xi}^{D}_{N,is}\, \frac{Q_i}{8\, m_i\, m_b}
\,\, ,
\label{c7A0h0}
\end{eqnarray}
where $m_i$ and $Q_i$ are the masses and charges of the down quarks 
($i=d,\,s,\,b$) respectively. 
These expressions show that the neutral Higgs bosons can give a large 
contribution to the coefficient $C_7$ which is in contradiction with 
the CLEO data  \cite{cleo2}, 
\begin{eqnarray}
Br (B\rightarrow X_s\gamma)= (3.15\pm 0.35\pm 0.32)\, 10^{-4} \,\, .
\label{br2}
\end{eqnarray}
Such dangerous terms can be removed by assuming that the couplings 
$\bar{\xi}^{D}_{N,is}$ ($i=d,s,b)$ and $\bar{\xi}^{D}_{N,db}$ are 
small enough to be able to reach the conditions 
$\bar{\xi}^{D}_{N,bb} \,\bar{\xi}^{D}_{N,is} <<1$ and 
$\bar{\xi}^{D}_{N,db} \,\bar{\xi}^{D}_{N,ds} <<1$.
The discussion given above results in the following restrictions:   
$\bar{\xi}^{D}_{N, ib} \sim 0$ and $\bar{\xi}^{D}_{N, ij}\sim 0$, where the 
indices $i,j$ denote d and s quarks . Further using the constraints 
\cite{alil1}, coming from the $\Delta F=2$ mixing, the $\rho$ parameter 
\cite{atwood}, and the measurement by CLEO Collaboration eq. (\ref{br2}) 
we get the condition  for $\bar{\xi}_{N, tc}$,  
$\bar{\xi}_{N, tc} << \bar{\xi}^{U}_{N, tt}$ and take into account only the 
Yukawa couplings of quarks $\bar{\xi}^{U}_{N,tt}$ and $\bar{\xi}^{D}_{N,bb}$. 
As for   $\bar{\xi}^{D}_{N,\tau\tau}$, we do 
not consider any constraint and increase this parameter to enhance the 
effects of neutral Higgs boson. 
(For further discussion about the restrictions of the model III parameters 
see \cite{atwood,alil1}.)

In this section, we study the Wilson coefficients $C_{Q_{1}}(m_{b})$ and
$C_{Q_{2}}(m_{b})$ coming from NHB effects and  $s$, 
$\frac{\bar{\xi}^{D}_{N,bb}}{m_b}$ and 
$\bar{\xi}^{D}_{N,\tau\tau}$ dependencies of $d\Gamma/ds$ and $\Gamma$ for 
the inclusive decay $b\rightarrow s \tau^+  \tau^-$, restricting 
$|C_7^{eff}|$ in the region $0.257 \leq |C_7^{eff}| \leq 0.439$
due to the CLEO measurement, eq.(\ref{br2}) (see \cite{alil1} for details). 
Our numerical calculations based on this restriction and throughout these 
calculations, we use the redefinition
\begin{eqnarray}
\xi^{U,D}=\sqrt{\frac{4 G_F}{\sqrt{2}}} \bar{\xi}^{U,D} \nonumber \,\, , 
\label{xineutr}
\end{eqnarray}
we take the scale $\mu=m_b$ and use the input values given in Table 
(\ref{input}). 
\newpage
\begin{table}[h]
        \begin{center}
        \begin{tabular}{|l|l|}
        \hline
        \multicolumn{1}{|c|}{Parameter} & 
                \multicolumn{1}{|c|}{Value}     \\
        \hline \hline
        $m_{\tau}$                   & $1.78$ (GeV) \\
        $m_c$                   & $1.4$ (GeV) \\
        $m_b$                   & $4.8$ (GeV) \\
        $\alpha_{em}^{-1}$      & 129           \\
        $\lambda_t$            & 0.04 \\
        Br ($B\rightarrow X_{c} \ell \bar{\nu}$)     & $0.103 \pm 0.01$\\
        $m_{t}$             & $175$ (GeV) \\
        $m_{W}$             & $80.26$ (GeV) \\
        $m_{Z}$             & $91.19$ (GeV) \\
        $\Lambda_{QCD}$             & $0.225$ (GeV) \\
        $\alpha_{s}(m_Z)$             & $0.117$  \\
        $sin\theta_W$             & $0.2325$  \\
        \hline
        \end{tabular}
        \end{center}
\caption{The values of the input parameters used in the numerical
          calculations.}
\label{input}
\end{table}

\begin{figure}[htb]
\vskip -3.0truein
\centering
\epsfxsize=6.8in
\leavevmode\epsffile{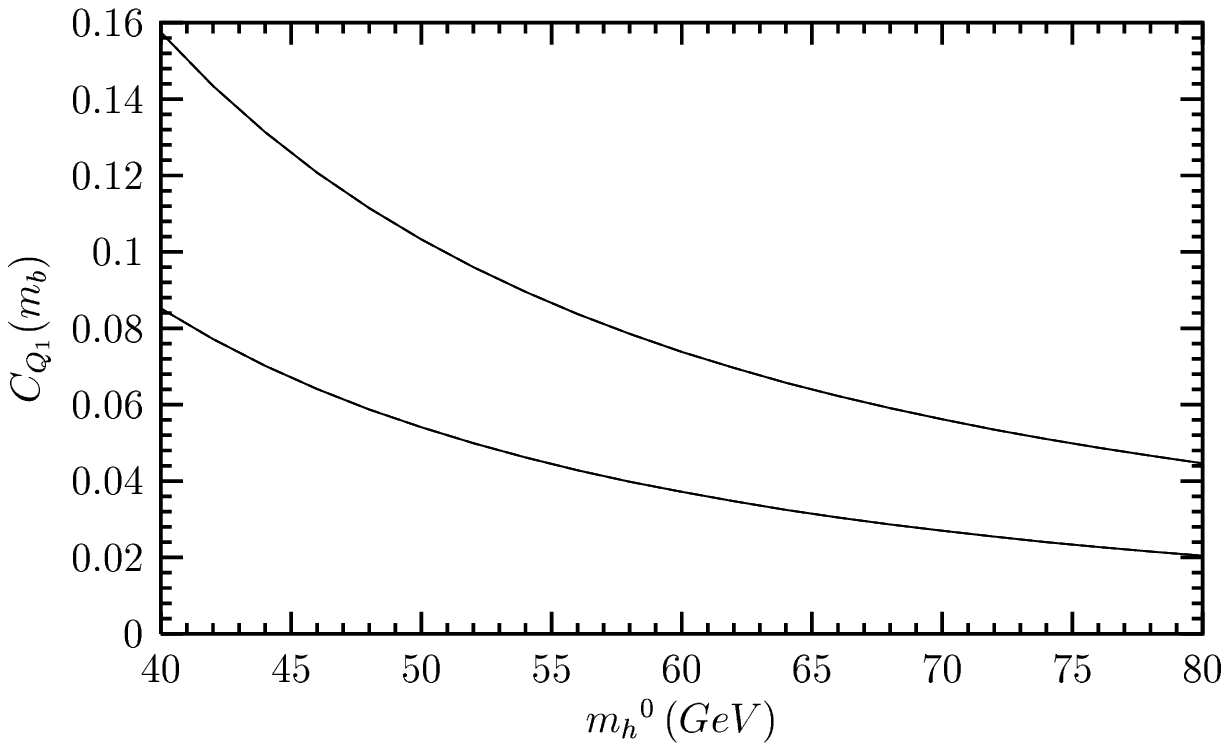}
\vskip -3.0truein
\caption[]{$C_{Q_{1}}(m_{b})$ as a function of  $m_{h^0}$
for  $\bar{\xi}_{N,bb}^{D}=40\, m_b$, $\bar{\xi}_{N,\tau\tau}^{D}=5 \, GeV$, 
$m_{H^{\pm}}=400\,GeV$ and $m_{H^0}=100\,GeV$ in case of the ratio $|r_{tb}|<1$.}
\label{CQ1mh0rtbK1}
\end{figure}

\begin{figure}[htb]
\vskip -3.0truein
\centering
\epsfxsize=6.8in
\leavevmode\epsffile{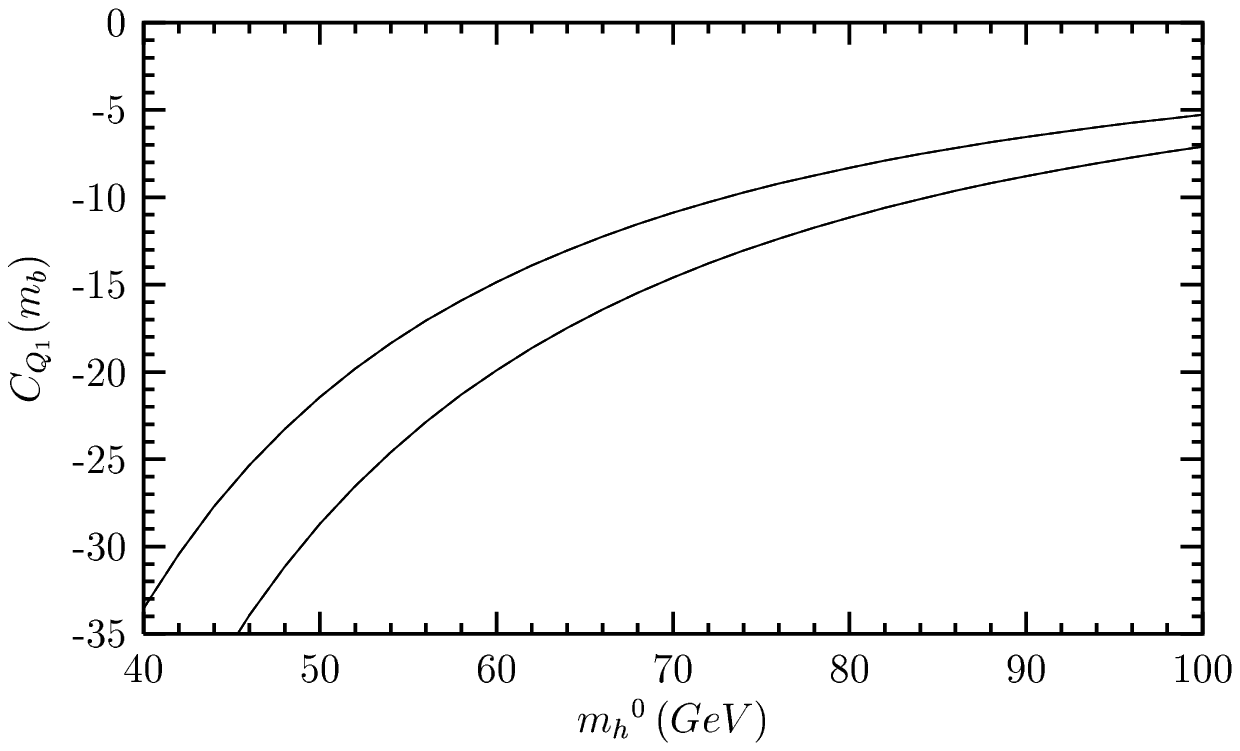}
\vskip -3.0truein
\caption[]{Same as Fig.\ref{CQ1mh0rtbK1}, but for $\bar{\xi}_{N,bb}^{D}=3\,
m_b$ and   $r_{tb}>1$.}
\label{CQ1mh0rtbB1}
\end{figure}

In Fig.~\ref{CQ1mh0rtbK1} (\ref{CQ1mh0rtbB1}), we present $m_{h^0}$
dependence of $C_{Q_{1}}(m_{b})$ for $C^{eff}_{7}>0$,
$\bar{\xi}_{N,bb}^{D}=40\, m_b$ ($3\, m_b$),
$\bar{\xi}_{N,\tau\tau}^{D}=5\, GeV$  in the case 
$|r_{tb}|=|\frac{\bar{\xi}_{N,tt}^{U}}{\bar{\xi}_{N,bb}^{D}}| <1 
\, (r_{tb}>1)$. Here $C_{Q_{1}}(m_{b})$ lies in the region 
bounded by solid lines. For $r_{tb}>1$, $m_{h^0}=80 \, GeV$ and $m_{H^0}=100
\, GeV$, the value of  
$C_{Q_{1}}(m_{b})$ changes between $0.020$ and  $0.045$. However for 
$r_{tb}>1$, we get values, -9 and -12, more than two orders of magnitude 
larger compared to  ones for $|r_{tb}|<1$, for the same value of $m_{h^{0}}$. 
Since $C_{Q_{1}}(m_{b})$ is 
directly proportional to $\bar{\xi}_{N,\tau\tau}^{D}$, its value may further 
increase with the increasing values of $\bar{\xi}_{N,\tau\tau}^{D}$. The 
corresponding 2HDM model II value of  $C_{Q_{1}}(m_{b})$ can be extracted 
from \cite{Chao} as beeing $\sim 0.4$  for large $\tan \beta$, 
$\tan \beta =25$. 

\begin{figure}[htb]
\vskip -3.0truein
\centering
\epsfxsize=6.8in
\leavevmode\epsffile{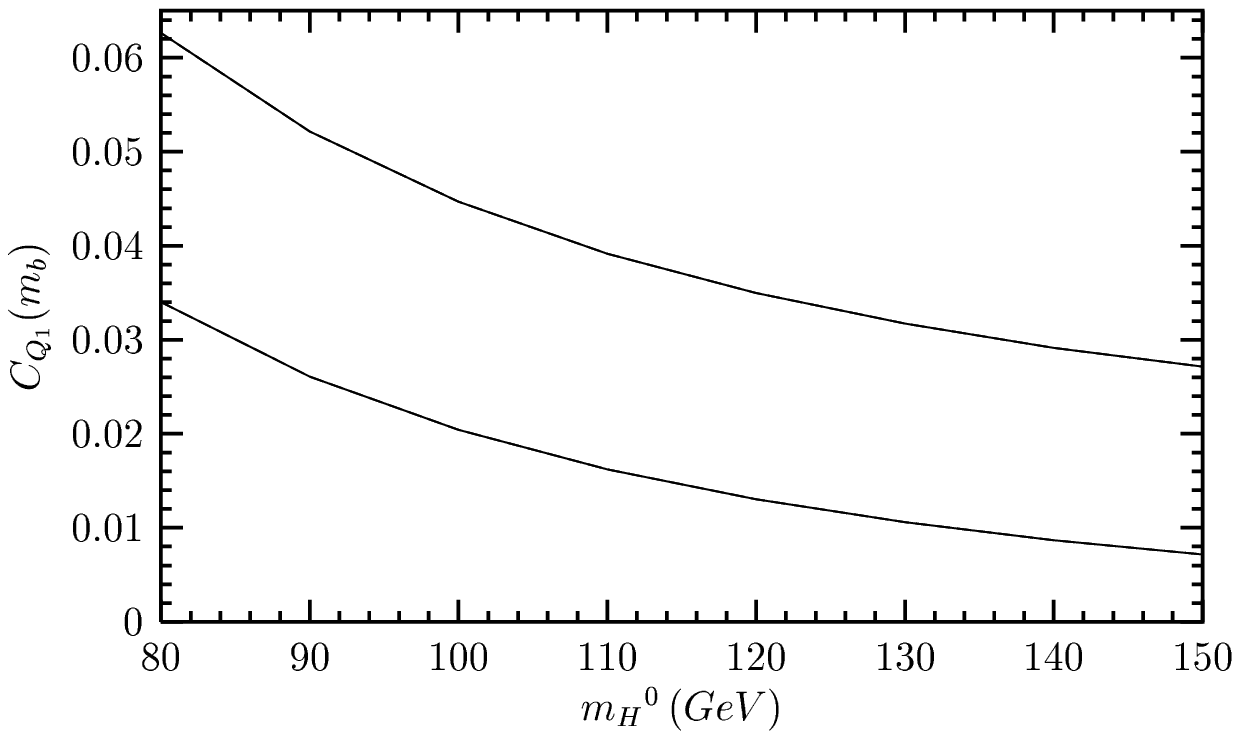}
\vskip -3.0truein
\caption[]{$C_{Q_{1}}(m_{b})$ as a function of  $m_{H^0}$
for  $\bar{\xi}_{N,bb}^{D}=40\, m_b$, $\bar{\xi}_{N,\tau\tau}^{D}=5 \, GeV$, 
$m_{H^{\pm}}=400\,GeV$ and $m_{h^0}=80\,GeV$ in case of the ratio 
$|r_{tb}|<1$.}
\label{CQ1mHH0rtbK1}
\end{figure}

\begin{figure}[htb]
\vskip -3.0truein
\centering
\epsfxsize=6.8in
\leavevmode\epsffile{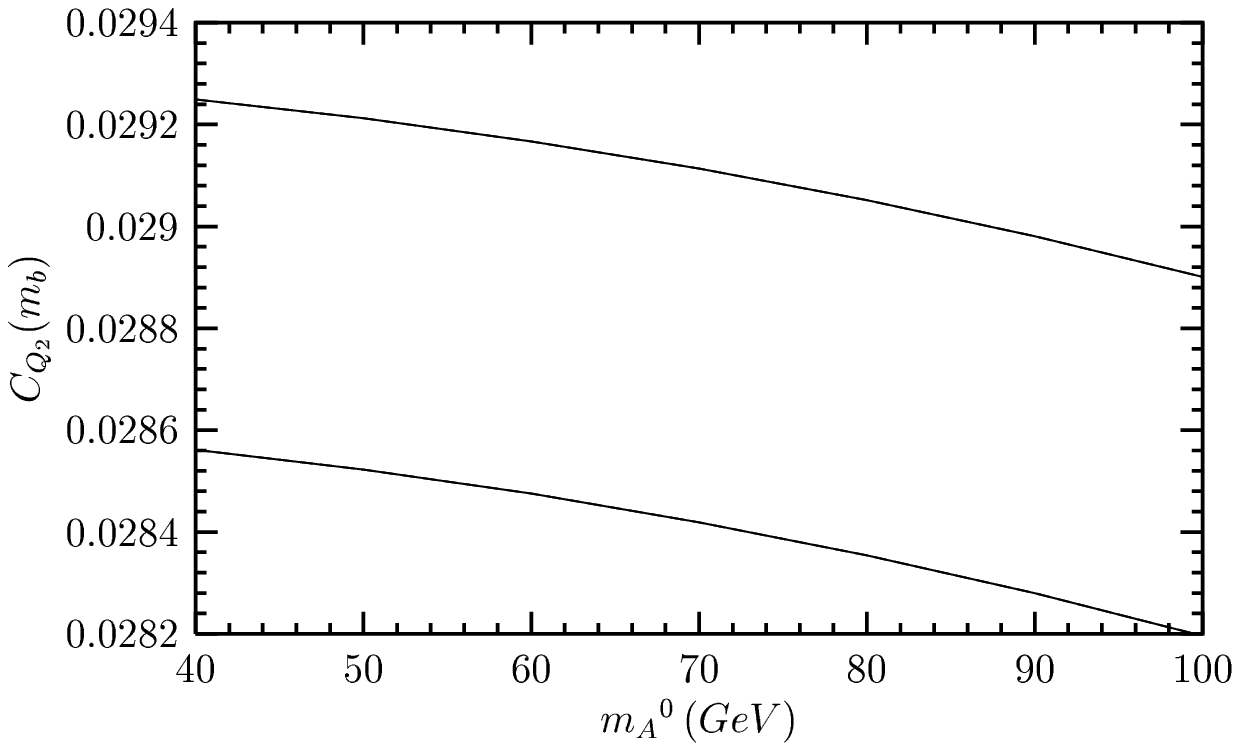}
\vskip -3.0truein
\caption[]{$C_{Q_{2}}(m_{b})$ as a function of  $m_{A^0}$
for  $\bar{\xi}_{N,bb}^{D}=40\, m_b$, $\bar{\xi}_{N,\tau\tau}^{D}=5 \, GeV$
and  $m_{H^{\pm}}=400\,GeV$ in case of the ratio $|r_{tb}|<1$.}
\label{CQ2mA0rtbK1}
\end{figure}

For completeness, in Figs.~\ref{CQ1mHH0rtbK1} and ~\ref{CQ2mA0rtbK1},  
we give $m_{H^0}$ dependence of $C_{Q_{1}}(m_{b})$ and $m_{A^0}$                                    dependence of $C_{Q_{2}}(m_{b})$, for $C^{eff}_{7}>0$,
$\bar{\xi}_{N,bb}^{D}=40\, m_b$, $\bar{\xi}_{N,\tau\tau}^{D}=5 \, GeV$  in the 
case $|r_{tb}|<1$. As seen from  
Fig.~\ref{CQ2mA0rtbK1}, $m_{A_0}$ dependence of $C_{Q_{2}}(m_{b})$ is 
relatively weaker and for $m_{A_0}=80 \, GeV$, $C_{Q_{2}}(m_{b})$ is between 
nearly  $0.0284$ and $0.0291$. For  $r_{tb}>1$, $C^{eff}_{7}>0$, 
$\bar{\xi}_{N,bb}^{D}=3\, m_b$ and $\bar{\xi}_{N,\tau\tau}^{D}=5 \, GeV$, 
$C_{Q_{2}}(m_{b})$ reaches up to the value of $-0.38$.  The 2HDM model II value of
$|C_{Q_{2}}(m_{b})|$ is $\sim 0.4$  for  $\tan \beta =25$ \cite{Chao}.    

Now we continue the analysis of the measurable quantities 
$\Gamma$ and $A_{FB}$ of the process under consideration. In the following, 
we use the numerical values $m_{H^{0}}=150\, GeV$, $m_{h^{0}}=80\, GeV$ 
and $m_{A^{0}}=80\, GeV$  in our  calculations.

\begin{figure}[htb]
\vskip -3.0truein
\centering
\epsfxsize=6.8in
\leavevmode\epsffile{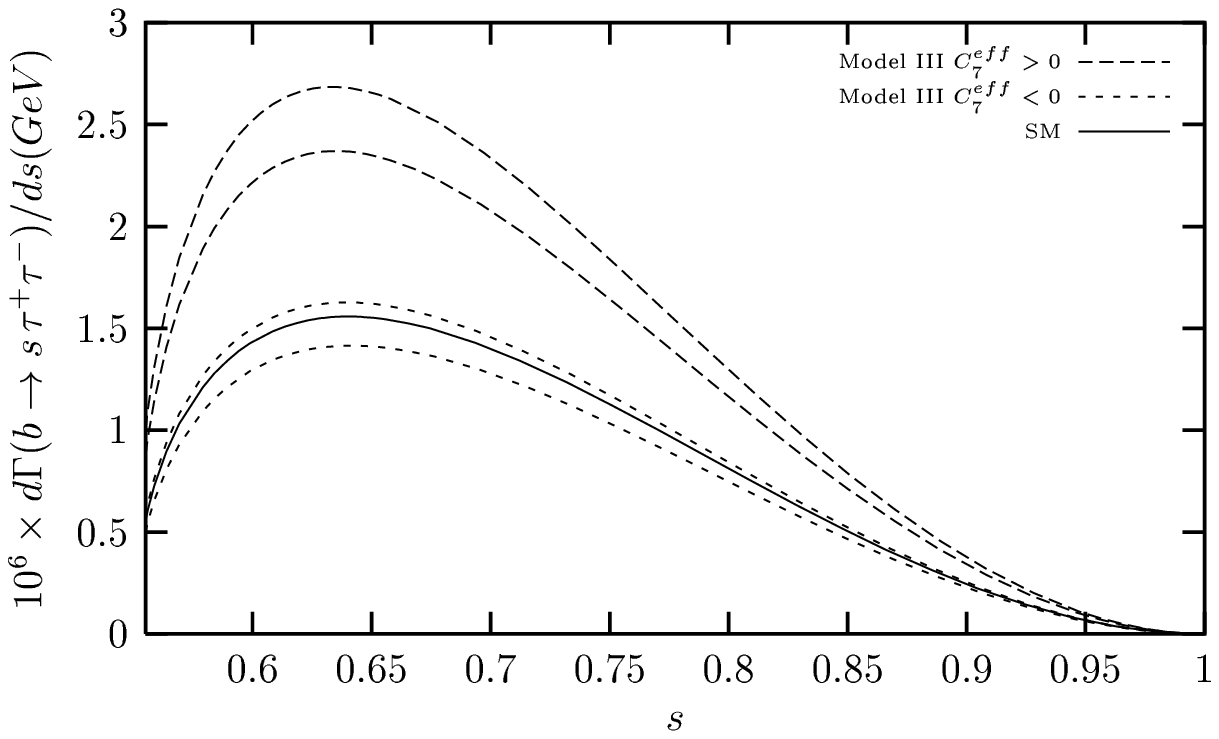}
\vskip -3.0truein
\caption[]{Differential $\Gamma$ as a function of  $s$ 
for  $\bar{\xi}_{N,bb}^{D}=40\, m_b$, $\bar{\xi}_{N,\tau\tau}^{D}=1 \, GeV$ 
and  $m_{H^{\pm}}=400\,GeV$ in case of the ratio $|r_{tb}|<1$.} 
\label{dGs0}
\end{figure}

\begin{figure}[htb]
\vskip -3.0truein
\centering
\epsfxsize=6.8in
\leavevmode\epsffile{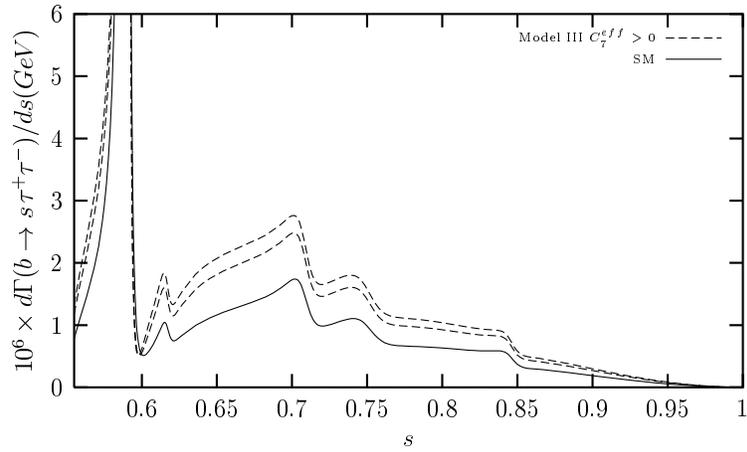}
\vskip -3.0truein
\caption[]{The same as Fig \ref{dGs0}, but with LD effects.}
\label{dGsLD}
\end{figure}

\begin{figure}[htb]
\vskip -3.0truein
\centering
\epsfxsize=6.8in
\leavevmode\epsffile{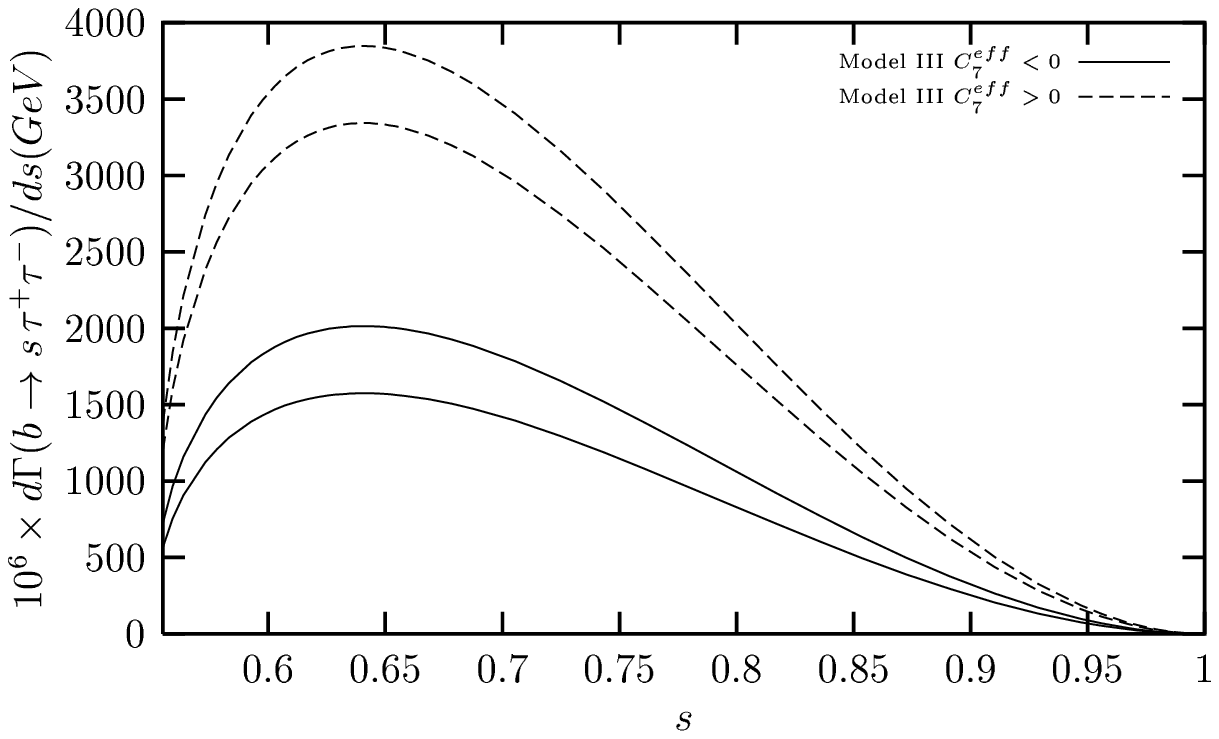}
\vskip -3.0truein
\caption[]{The same as Fig \ref{dGsLD}, but for $\bar{\xi}_{N,bb}^{D}=3\,
m_b$ and $r_{tb}>1$.}
\label{dGsLD2}
\end{figure}

In Fig.~\ref{dGs0}, we plot the differential $\Gamma$ of the decay 
$b\rightarrow s \tau^+ \tau^-$ with respect to the parameter $s$ for 
$\bar{\xi}_{N,bb}^{D}=40\, m_b$, 
$\bar{\xi}_{N,\tau\tau}^{D}=1 \, GeV$ 
and charged Higgs mass $m_{H^{\pm}}=400\, GeV$ in case of the ratio
$|r_{tb}|<1$. Here the differential $\Gamma$ lies in the region bounded 
by dashed (small dashed) curves for $C_7^{eff} > 0$ ($C_7^{eff} < 0$). A 
small enhancement is possible especially for $C_7^{eff} > 0$ case compared 
to the SM (solid curve). Further, the restriction region of the differential 
$\Gamma$ for model III becomes narrower with increasing or decreasing 
values of the parameter $s$. Fig. \ref{dGsLD} is devoted the same dependence 
of the differential $\Gamma$ including the long distance (LD) effects. 
Here $C_7^{eff} < 0$ case for model III almost coincides with the SM 
(solid curve). In case of the ratio $r_{tb} >1$, extremely large 
enhancement, 3 orders larger compared the $|r_{tb}| <1$ case, is 
reached even for the small values  of $\bar{\xi}_{N,bb}^{D}$ 
(see Fig.~\ref{dGsLD2}).  

\begin{figure}[htb]
\vskip -3.0truein
\centering
\epsfxsize=6.8in
\leavevmode\epsffile{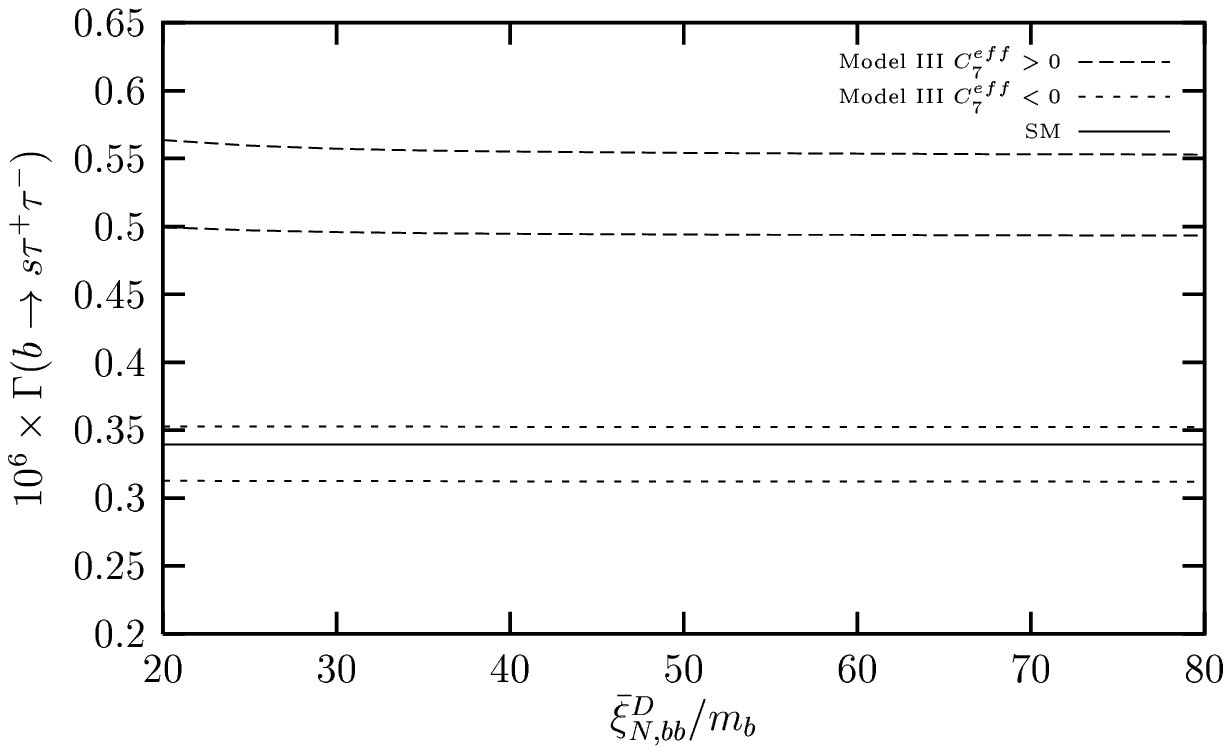}
\vskip -3.0truein
\caption[]{$\Gamma$ as a function of  $\frac{\bar{\xi}_{N,bb}^{D}}{m_b}$ 
for $\bar{\xi}_{N,\tau\tau}^{D}=1 \, GeV$  and  $m_{H^{\pm}}=400\,GeV$ 
in case of the ratio $|r_{tb}|<1$.}
\label{GamaLDbb}
\end{figure}

\begin{figure}[htb] 
\vskip -3.0truein   
\centering
\epsfxsize=6.8in    
\leavevmode\epsffile{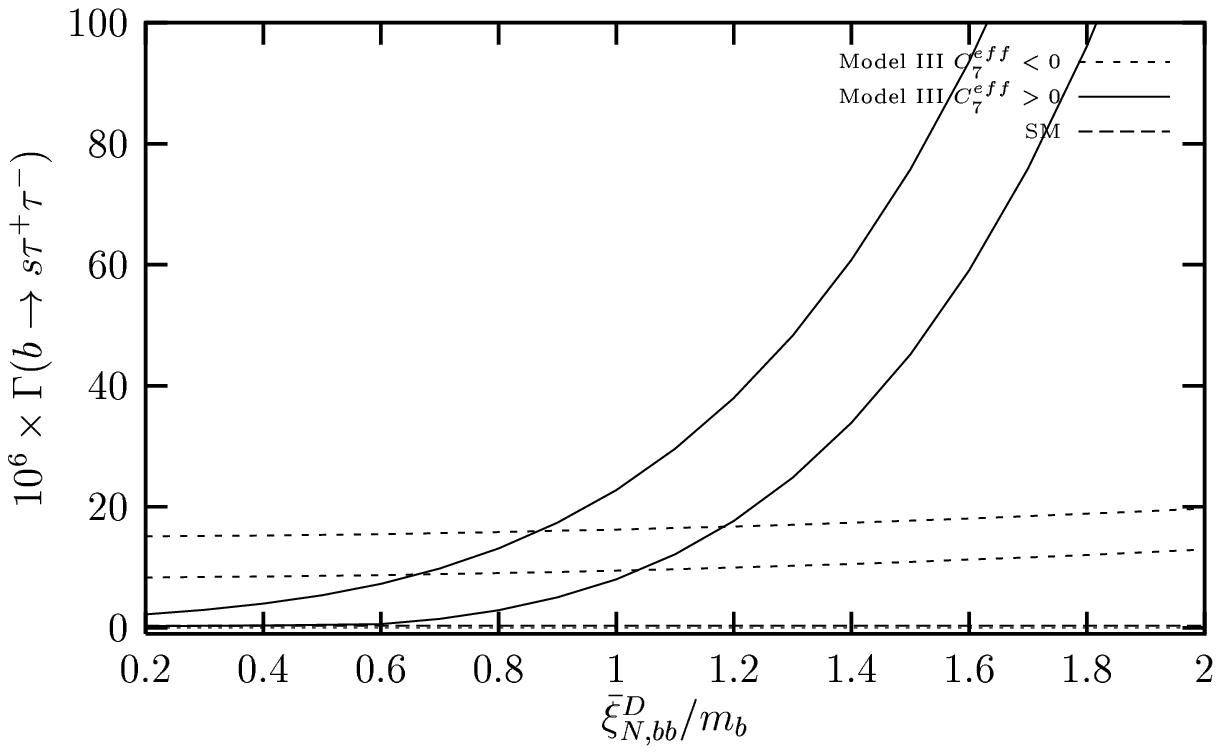}
\vskip -3.0truein   
\caption[]{The same as Fig. \ref{GamaLDbb} but for $r_{tb}>1$.} 
\label{GamaLDbb2}    
\end{figure}

Fig.~\ref{GamaLDbb} shows $\frac{\bar{\xi}_{N,bb}^{D}}{m_b}$ dependence of 
$\Gamma$ of the decay under consideration for
$\bar{\xi}_{N,\tau\tau}^{D}=1 \, GeV$ 
and charged Higgs mass $m_{H^{\pm}}=400\, GeV$ in case of the ratio
$|r_{tb}|<1$. Here  $\Gamma$ is almost non-sensitive to 
$\frac{\bar{\xi}_{N,bb}^{D}}{m_b}$. However for $r_{tb}>1$ case 
(Fig. \ref{GamaLDbb2}) 
$\Gamma$ is strongly sensitive to $\frac{\bar{\xi}_{N,bb}^{D}}{m_b}$ for 
$C_7^{eff} > 0$. Further, $\Gamma$ is 2 orders (3 orders) larger compared 
to the SM result for $C_7^{eff} < 0$ ($C_7^{eff} > 0$) even for 
$\frac{\bar{\xi}_{N,bb}^{D}}{m_b} < 2$ .

\begin{figure}[htb]
\vskip -3.0truein
\centering
\epsfxsize=6.8in
\leavevmode\epsffile{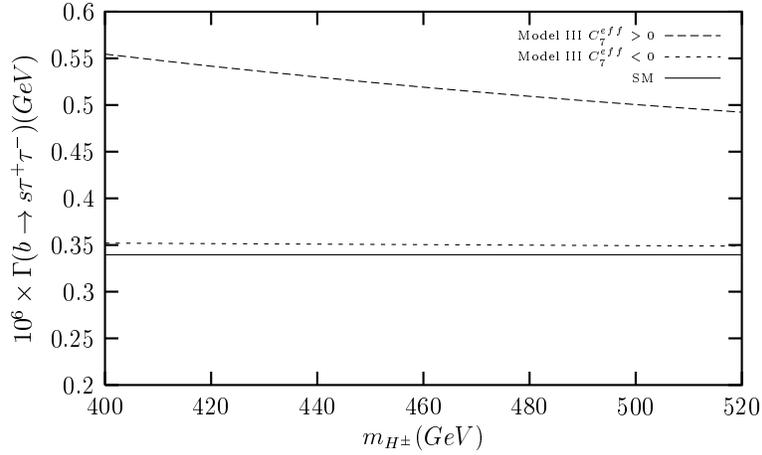}
\vskip -3.0truein
\caption[]{$\Gamma$ as a function of  $m_{H^{\pm}}$ for
$\bar{\xi}_{N,bb}^{D}=40\, m_b$, $\bar{\xi}_{N,\tau\tau}^{D}=1 \, GeV $
in case of the ratio $|r_{tb}|<1$.}
\label{GamaLDmh} 
\end{figure}

Fig.~\ref{GamaLDmh} is devoted to the dependence of $\Gamma$ to the 
charged Higgs mass $m_H^{\pm}$. $\Gamma$ has a weak dependence (almost 
no dependence) on $m_H^{\pm}$ for $C_7^{eff} > 0$ ($C_7^{eff} < 0$). 

\begin{figure}[htb]
\vskip -3.0truein
\centering
\epsfxsize=6.8in
\leavevmode\epsffile{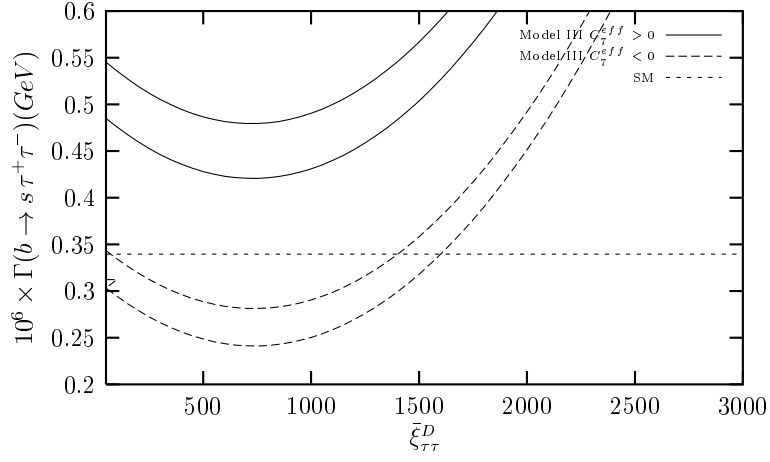}
\vskip -3.0truein
\caption[]{$\Gamma$ as a function of $\bar{\xi}_{N,\tau\tau}^{D}$ ,
for $\bar{\xi}_{N,bb}^{D}=40\, m_b$, $m_{H^{\pm}}=400\, GeV$, in case of 
the ratio $|r_{tb}|<1$. }
\label{GamaLDtau}
\end{figure}

\begin{figure}[htb]
\vskip -3.0truein
\centering
\epsfxsize=6.8in
\leavevmode\epsffile{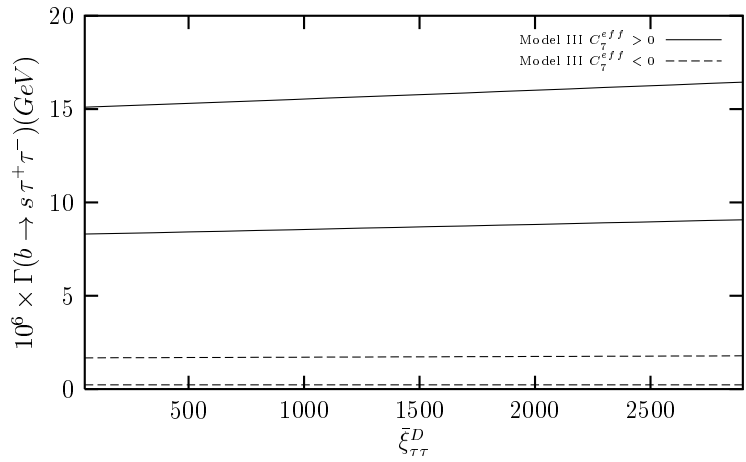}
\vskip -3.0truein
\caption[]{The same as Fig. \ref{GamaLDtau} for $\bar{\xi}_{N,bb}^{D}=0.1\,
m_b$ in case of the ratio $r_{tb}>1$. }
\label{GamaLDtau2}         
\end{figure}

For completeness, in Figs. ~\ref{GamaLDtau} and ~\ref{GamaLDtau2} we also 
present $\bar{\xi}_{N,\tau\tau}^{D}$ dependence of $\Gamma$ for large 
values of $\bar{\xi}_{N,\tau\tau}^{D}$. Sensitivity of $\Gamma$ to 
$\bar{\xi}_{N,\tau\tau}^{D}$ increases with the increasing values of this
parameter. $\Gamma$ enhances for extremely large values of 
$\bar{\xi}_{N,\tau\tau}^{D}$ and this is the contribution due to the NHB
effects. For $|r_{tb}|<1$ the NHB effects are small and destructive up to 
the large values of $\bar{\xi}_{N,\tau\tau}^{D}$, 
$\bar{\xi}_{N,\tau\tau}^{D}=800 \, GeV$.  For $C_7^{eff} > 0$,
$\frac{\bar{\xi}_{N,bb}^{D}}{m_b}=40$ and 
$\bar{\xi}_{N,\tau\tau}^{D}=1 (100) \, GeV$ this effect is at the 
order of the magnitude $\%\, 0.1\,(4)$ of the overall contribution. However, 
it is positive for $r_{tb}>1$ and it becomes considerable with 
increasing values of $\bar{\xi}_{N,\tau\tau}^{D}$. For $C_7^{eff} > 0$, 
the small value $\frac{\bar{\xi}_{N,bb}^{D}}{m_b}=3$ and 
$\bar{\xi}_{N,\tau\tau}^{D}=1 (100,\, 200) \, GeV$, the NHB 
contribution can reach the magnitude $\%\, 0.15\,(7, \,26)$ of the overall 
contribution.

\begin{figure}[htb]
\vskip -3.0truein
\centering
\epsfxsize=6.8in
\leavevmode\epsffile{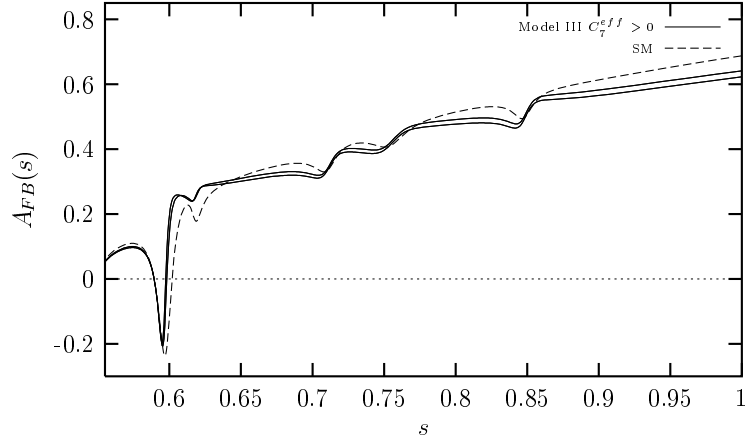}
\vskip -3.0truein
\caption[]{Differential $A_{FB}$ as a function of $s$ 
for $\bar{\xi}_{N,bb}^{D}=40\, m_b$, $\bar{\xi}_{N,\tau\tau}^{D}=1 \, GeV$
and  $m_{H^{\pm}}=400\, GeV$ including LD effects in case of the ratio
$|r_{tb}| <1$.}
\label{dAsmLD}
\end{figure}

\begin{figure}[htb]
\vskip -3.0truein
\centering
\epsfxsize=6.8in
\leavevmode\epsffile{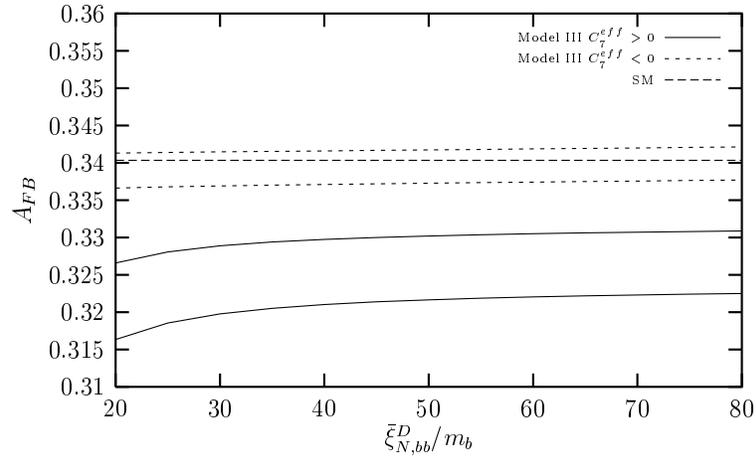}
\vskip -3.0truein
\caption[]{ $A_{FB}$ as a function of
$\frac{\bar{\xi}_{N,bb}^{D}}{m_b}$ for $\bar{\xi}_{N,\tau\tau}^{D}=1 \, GeV$,
$m_{H^{\pm}}=400\, GeV$ and $|r_{tb}|<1$.}
\label{AsmLDbb}
\end{figure}

\begin{figure}[htb]
\vskip -3.0truein
\centering
\epsfxsize=6.8in
\leavevmode\epsffile{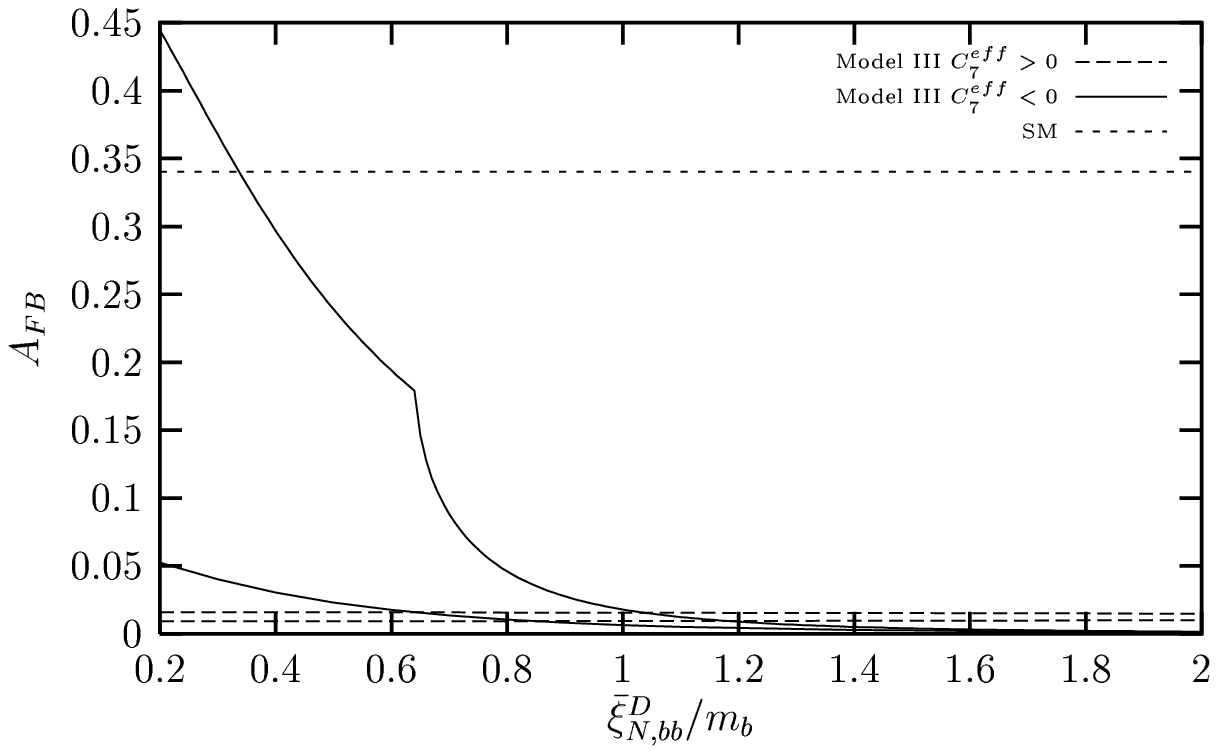}
\vskip -3.0truein
\caption[]{ The same as Fig. \ref{AsmLDbb} but for $r_{tb}>1$.}
\label{AsmLDbb2} 
\end{figure}

Our results on  $A_{FB}(s)$ and  $A_{FB}$ for the decay under 
consideration are presented through the graphs given by 
Figs. ~\ref{dAsmLD}-~\ref{AsmLDbb2}. In Fig.~\ref{dAsmLD}  
$A_{FB} (s)$ is 
shown for $\bar{\xi}_{N,bb}^{D}=40\, m_b$, 
$\bar{\xi}_{N,\tau\tau}^{D}=1 \, GeV$ 
and charged Higgs mass $m_{H^{\pm}}=400\, GeV$ in case of the ratio
$|r_{tb}| <1$. Here $A_{FB} (s)$ lies in the region bounded 
by solid  lines for $C_7^{eff} > 0$. Dashed line presents $C_7^{eff}< 0$
case and the SM result coincides with this line. There is possible negative
values of $A_{FB} (s)$ due to the LD effects. For $r_{tb} >1$, 
$A_{FB} (s)$ almost vanishes ($\sim\, 10^{-4}$).

Fig. \ref{AsmLDbb} is devoted to $\frac{\bar{\xi}_{N,bb}^{D}}{m_b}$ 
dependence of $A_{FB}$  for $\bar{\xi}_{N,\tau\tau}^{D}=1 \, GeV$, 
charged Higgs mass $m_{H^{\pm}}=400\, GeV$ and $|r_{tb}|<1$. Here, $A_{FB}$ 
is not sensitive to $\frac{\bar{\xi}_{N,bb}^{D}}{m_b}$, especially for large 
values of this parameter. The SM and model III average results for 
$C_7^{eff}< 0$ ($C_7^{eff}> 0$) are $0.340$ and $0.340$ ($0.325$), 
respectively. $A_{FB}$ is sensitive to the parameter 
$\frac{\bar{\xi}_{N,bb}^{D}}{m_b}$ for its small values in the case where 
$r_{tb}>1$ and $C_7^{eff}< 0$ (Fig. \ref{AsmLDbb2}). The enhancement over the
SM is possible for $\frac{\bar{\xi}_{N,bb}^{D}}{m_b} <0.4$, namely $A_{FB}$ can
reach the value of $0.45$. The restriction region for $A_{FB}$ is large for this 
case. However, for $C_7^{eff}> 0$, $A_{FB}$ almost vanishes. 
 
The NHB effects on $A_{FB}$ is sensitive to the coupling 
$\bar{\xi}_{N,\tau\tau}^{D}$ as it should be. For $|r_{tb}|<1$ and 
$C_7^{eff}> 0$, the NHB contribution is $-\%\, 0.15$ for 
$\bar{\xi}_{N,\tau\tau}^{D}=1 \, GeV$ and $\%\, 1.2$ for 
$\bar{\xi}_{N,\tau\tau}^{D}=100 \, GeV$ in case the parameter 
$\frac{\bar{\xi}_{N,bb}^{D}}{m_b}=40$. Increasing 
$\bar{\xi}_{N,\tau\tau}^{D}$ causes to the enhancement in the NHB effects. 
For $r_{tb}>1$ and $C_7^{eff}< 0$, the NHB effects are negative and it 
increases the overall result by  $\%\, 10$ for 
$\frac{\bar{\xi}_{N,bb}^{D}}{m_b} = 0.4$ and 
$\bar{\xi}_{N,\tau\tau}^{D}=100 \, GeV$. For $r_{tb}>1$ and 
$C_7^{eff}> 0$, the NHB effects to $A_{FB}$ are negligible. 

Now, we would like to summarize our results. 
\begin{itemize}
\item $\Gamma$ for the process under consideration is at the order of $
10^{-6}$ for $|r_{tb}| <1$ and $C_{7}^{eff}>0$ results is greater compared 
to $C_{7}^{eff}<0$ one. On the otherhand, for $r_{tb}> 1$, there is a
considerable enhancement, three order larger compared to the SM case even
for small values of $\frac{\bar{\xi}_{N,bb}^{D}}{m_b}$. Further, $\Gamma$ 
is not sensitive to $\frac{\bar{\xi}_{N,bb}^{D}}{m_b}$ for $|r_{tb}| <1$, 
however strong sensitivity to this parameter is observed for $r_{tb} >1$.

\item $A_{FB}$ is not so much sensitive to the model III parameters for 
$|r_{tb}| <1$. For $r_{tb} >1$ , there is a possible enhancement in the 
$A_{FB}$ for small values of $\frac{\bar{\xi}_{N,bb}^{D}}{m_b}$, however 
it becomes negligible with increasing $\frac{\bar{\xi}_{N,bb}^{D}}{m_b}$. 

\item The NHB effects becomes important for the large values of the Yukawa
coupling $\bar{\xi}_{N,\tau\tau}^{D}$.

\end{itemize}

Therefore, the experimental investigation of $\Gamma$ and $A_{FB}$ ensure 
a crucial test for new physics and also the sign of $C_{7}^{eff}$.

\newpage
{\bf \LARGE {Appendix}} \\
\begin{appendix}
\section{The operator basis }
The operator basis in the  2HDM (model III ) for our process  
is \cite{Chao,Grinstein2,misiak}
\begin{eqnarray}
 O_1 &=& (\bar{s}_{L \alpha} \gamma_\mu c_{L \beta})
               (\bar{c}_{L \beta} \gamma^\mu b_{L \alpha}), \nonumber   \\
 O_2 &=& (\bar{s}_{L \alpha} \gamma_\mu c_{L \alpha})
               (\bar{c}_{L \beta} \gamma^\mu b_{L \beta}),  \nonumber   \\
 O_3 &=& (\bar{s}_{L \alpha} \gamma_\mu b_{L \alpha})
               \sum_{q=u,d,s,c,b}
               (\bar{q}_{L \beta} \gamma^\mu q_{L \beta}),  \nonumber   \\
 O_4 &=& (\bar{s}_{L \alpha} \gamma_\mu b_{L \beta})
                \sum_{q=u,d,s,c,b}
               (\bar{q}_{L \beta} \gamma^\mu q_{L \alpha}),   \nonumber  \\
 O_5 &=& (\bar{s}_{L \alpha} \gamma_\mu b_{L \alpha})
               \sum_{q=u,d,s,c,b}
               (\bar{q}_{R \beta} \gamma^\mu q_{R \beta}),   \nonumber  \\
 O_6 &=& (\bar{s}_{L \alpha} \gamma_\mu b_{L \beta})
                \sum_{q=u,d,s,c,b}
               (\bar{q}_{R \beta} \gamma^\mu q_{R \alpha}),  \nonumber   \\  
 O_7 &=& \frac{e}{16 \pi^2}
          \bar{s}_{\alpha} \sigma_{\mu \nu} (m_b R + m_s L) b_{\alpha}
                {\cal{F}}^{\mu \nu},                             \nonumber  \\
 O_8 &=& \frac{g}{16 \pi^2}
    \bar{s}_{\alpha} T_{\alpha \beta}^a \sigma_{\mu \nu} (m_b R +
m_s L)  
          b_{\beta} {\cal{G}}^{a \mu \nu} \nonumber \,\, , \\  
 O_9 &=& \frac{e}{16 \pi^2}
          (\bar{s}_{L \alpha} \gamma_\mu b_{L \alpha})
              (\bar{\tau} \gamma^\mu \tau)  \,\, ,    \nonumber    \\
 O_{10} &=& \frac{e}{16 \pi^2}
          (\bar{s}_{L \alpha} \gamma_\mu b_{L \alpha})
              (\bar{\tau} \gamma^\mu \gamma_{5} \tau)  \,\, ,    \nonumber  \\
Q_1&=&   \frac{e^2}{16 \pi^2}(\bar{s}^{\alpha}_{L}\,b^{\alpha}_{R})\,(\bar{\tau}\tau ) 
\nnb  \\ 
Q_2&=&    \frac{e^2}{16 \pi^2}(\bar{s}^{\alpha}_{L}\,b^{\alpha}_{R})\,
(\bar{\tau} \gamma_5 \tau ) \nnb \\
Q_3&=&    \frac{g^2}{16 \pi^2}(\bar{s}^{\alpha}_{L}\,b^{\alpha}_{R})\,
\sum_{q=u,d,s,c,b }(\bar{q}^{\beta}_{L} \, q^{\beta}_{R} ) \nnb \\
Q_4&=&  \frac{g^2}{16 \pi^2}(\bar{s}^{\alpha}_{L}\,b^{\alpha}_{R})\,
\sum_{q=u,d,s,c,b } (\bar{q}^{\beta}_{R} \, q^{\beta}_{L} ) \nnb \\
Q_5&=&   \frac{g^2}{16 \pi^2}(\bar{s}^{\alpha}_{L}\,b^{\beta}_{R})\,
\sum_{q=u,d,s,c,b } (\bar{q}^{\beta}_{L} \, q^{\alpha}_{R} ) \nnb \\
Q_6&=&   \frac{g^2}{16 \pi^2}(\bar{s}^{\alpha}_{L}\,b^{\beta}_{R})\,
\sum_{q=u,d,s,c,b } (\bar{q}^{\beta}_{R} \, q^{\alpha}_{L} ) \nnb \\
Q_7&=&   \frac{g^2}{16 \pi^2}(\bar{s}^{\alpha}_{L}\,\sigma^{\mu \nu} \, 
b^{\alpha}_{R})\,
\sum_{q=u,d,s,c,b } (\bar{q}^{\beta}_{L} \, \sigma_{\mu \nu } 
q^{\beta}_{R} ) \nnb \\
Q_8&=&    \frac{g^2}{16 \pi^2}(\bar{s}^{\alpha}_{L}\,\sigma^{\mu \nu} 
\, b^{\alpha}_{R})\,
\sum_{q=u,d,s,c,b } (\bar{q}^{\beta}_{R} \, \sigma_{\mu \nu } 
q^{\beta}_{L} ) \nnb \\ 
Q_9&=&   \frac{g^2}{16 \pi^2}(\bar{s}^{\alpha}_{L}\,\sigma^{\mu \nu} 
\, b^{\beta}_{R})\,
\sum_{q=u,d,s,c,b }(\bar{q}^{\beta}_{L} \, \sigma_{\mu \nu } 
q^{\alpha}_{R} ) \nnb \\
Q_{10}&= & \frac{g^2}{16 \pi^2}(\bar{s}^{\alpha}_{L}\,\sigma^{\mu \nu} \, 
b^{\beta}_{R})\,
\sum_{q=u,d,s,c,b }(\bar{q}^{\beta}_{R} \, \sigma_{\mu \nu } q^{\alpha}_{L} )
\label{op1}
\end{eqnarray}
where $\alpha$ and $\beta$ are $SU(3)$ colour indices and 
${\cal{F}}^{\mu \nu}$ and ${\cal{G}}^{\mu \nu}$ are the field strength 
tensors of the electromagnetic and strong interactions, respectively. Note 
that there are also flipped chirality partners of these operators, which 
can be obtained by interchanging $L$ and $R$ in the basis given above in 
model III. However, we do not present them here since corresponding  Wilson 
coefficients are negligible.
\section{The Initial values of the Wilson coefficients.}
The initial values of the Wilson coefficients for the relevant process 
in the SM are \cite{Grinstein2}
\begin{eqnarray}
C^{SM}_{1,3,\dots 6}(m_W)&=&0 \nonumber \, \, , \\
C^{SM}_2(m_W)&=&1 \nonumber \, \, , \\
C_7^{SM}(m_W)&=&\frac{3 x_t^3-2 x_t^2}{4(x_t-1)^4} \ln x_t+
\frac{-8 x_t^3-5 x_t^2+7 x_t}{24 (x_t-1)^3} \nonumber \, \, , \\
C_8^{SM}(m_W)&=&-\frac{3 x_t^2}{4(x_t-1)^4} \ln x_t+
\frac{-x_t^3+5 x_t^2+2 x_t}{8 (x_t-1)^3}\nonumber \, \, , \\ 
C_9^{SM}(m_W)&=&-\frac{1}{sin^2\theta_{W}} B(x_t) +
\frac{1-4 \sin^2 \theta_W}{\sin^2 \theta_W} C(x_t)-D(x_t)+
\frac{4}{9}, \nonumber \, \, , \\
C_{10}^{SM}(m_W)&=&\frac{1}{\sin^2\theta_W}
(B(x_t)-C(x_t))\nonumber \,\, , \\
C_{Q_i}^{SM}(m_W) & = & 0~~~ i=1,..,10~.
\end{eqnarray}
The initial values for the additional part due to charged Higgs bosons are 
\begin{eqnarray}
C^{H}_{1,\dots 6 }(m_W)&=&0 \nonumber \, , \\
C_7^{H}(m_W)&=& Y^2 \, F_{1}(y_t)\, + \, X Y \,  F_{2}(y_t) 
\nonumber  \, \, , \\
C_8^{H}(m_W)&=& Y^2 \,  G_{1}(y_t) \, + \, X Y \, G_{2}(y_t) 
\nonumber\, \, , \\
C_9^{H}(m_W)&=&  Y^2 \,  H_{1}(y_t) \nonumber  \, \, , \\
C_{10}^{H}(m_W)&=& Y^2 \,  L_{1}(y_t)  
\label{CH} \, \, , 
\end{eqnarray}
where 
\bea
X & = & \frac{1}{m_{b}}~~~\left(\bar{\xi}^{D}_{N,bb}+\bar{\xi}^{D}_{N,sb}
\frac{V_{ts}}{V_{tb}} \right) ~~,~~ \nnb \\
Y & = & \frac{1}{m_{t}}~~~\left(\bar{\xi}^{U}_{N,tt}+\bar{\xi}^{U}_{N,tc}
\frac{V^{*}_{cs}}{V^{*}_{ts}} \right) ~~,~~
\eea
and due to the neutral  Higgs bosons are 
\bea
\!\!\!\!\!\!\!\!\!\!\!\!\!\!\!\!\!\!\!\!\!\!\!\!\!\!\!\!\!\!\!\!\!\!\!\!\!\!\!
\!\!\!\!\!\!\!\!\!\!\!\!\!\!\!\!\!\!\!\!\!\!\!\!\!\!\!\!\!\!\!\!\!\!\!\!\!\!\!
C^{A^{0}}_{Q_{2}}((\bar{\xi}^{U}_{N,tt})^{3}) =
\frac{\bar{\xi}^{D}_{N,\tau \tau}(\bar{\xi}^{U}_{N,tt})^{3}
m_{b} y_t (\Theta_5 (y_t)z_A-\Theta_1 (z_{A},y_t))}{32 \pi^{2}m_{A^{0}}^{2} 
m_{t} \Theta_1 (z_{A},y_t) \Theta_5 (y_t)} , \nnb
\eea
\bea
C^{A^{0}}_{Q_{2}}((\bar{\xi}^{U}_{N,tt})^{2})=\frac{\bar{\xi}^{D}_{N,\tau\tau}
(\bar{\xi}^{U}_{N,tt})^{2}
\bar{\xi}^{D}_{N,bb}}{32 \pi^{2}  m_{A^{0}}^{2}}\Big{(} 
\frac{(y_t (\Theta_1 (z_{A},y_t) - \Theta_5 (y_t) (xy+z_A))-
2 \Theta_1 (z_{A},y_t) \Theta_5 (y_t)   \ln [\frac{z_A \Theta_5 (y_t)}
{ \Theta_1 (z_{A},y_t)}]}{ \Theta_1 (z_{A},y_t) \Theta_5 (y_t)}\Big{)}, \nnb
\eea
\bea
C^{A^{0}}_{Q_{2}}(\bar{\xi}^{U}_{N,tt}) &=& 
\frac{g^2\bar{\xi}^{D}_{N,\tau\tau}\bar{\xi}^{U}_{N,tt} m_b
x_t}{64 \pi^2 m_{A^{0}}^{2}  m_t } \Bigg{(}\frac{2}{\Theta_5 (x_t)}
- \frac{xy x_t+2 z_A}{\Theta_1 (z_{A},x_t)}-2
\ln [\frac{z_A \Theta_5(x_t)}{ \Theta_1 (z_{A},x_t)}]\nnb \\ &
&\!\!\!\!\!\!\!\!\!\!\!\!\!\!\!\!\!\!\!\!\!\!\!\!\!\!\!\!\!\!\!\!\!\!\!\!\!\!
- x y x_t y_t \Big{(}\frac{(x-1) x_t 
(y_t/z_A-1)-(1+x)y_t}{(\Theta_6 -(x-y)(x_t -y_t))(\Theta_3
(z_A)+(x-y)(x_t-y_t)z_A)}-
\frac{x (y_t+x_t(1-y_t/z_A))-2 y_t }{\Theta_6 \Theta_3 (z_A)}\Big{)}
\Bigg{)}\, , \nnb
\eea
\bea
\!\!\!C^{A^{0}}_{Q_{2}}(\bar{\xi}^{D}_{N,bb}) =
\frac{g^2\bar{\xi}^{D}_{N,\tau\tau}\bar{\xi}^{D}_{N,bb}}{64 \pi^2 m^2_{A^{0}} }
\Big{(}1-
\frac{x^2_t y_t+2 y (x-1)x_t y_t-z_A (x^2_t+\Theta_6)}{ \Theta_3 (z_A)}+
\frac{x^2_t (1-y_t/z_A)}{\Theta_6}+2 \ln [\frac{z_A \Theta_6}{ \Theta_2
(z_{A})}]
\Big{)}\, ,\nnb 
\eea
\bea
\lefteqn{\!\!\!\!\!\!\!\!\!\!\!\!\!\!\!\!\!\!\!\!\!C^{H^{0}}_{Q_{1}}
((\bar{\xi}^{U}_{N,tt})^{2}) = 
\frac{g^2 (\bar{\xi}^{U}_{N,tt})^2 m_b m_{\tau}
}{64 \pi^2 m^2_{H^{0}} m^2_t } \Bigg{(}
\frac{x_t (1-2 y) y_t}{\Theta_5 (y_t)}+\frac{(-1+2 \cos^2 \theta_W) (-1+x+y) 
y_t} {\cos^2 \theta_W \Theta_4 (y_t)} } \nnb
\\ & &
+\frac{z_H (\Theta_1 (z_H,y_t) x y_t + 
\cos^2 \theta_W \,(-2 x^2 (-1+x_t) y y^2_t+x x_t y y^2_t-\Theta_8 z_H))}
{\cos^2 \theta_W \Theta_1 (z_H,y_t) \Theta_7 }\Bigg{)} ,             
\label{NHB}
\eea
\bea
\!\!\!\!\!\!\!\!\!\!\!\!\!\!\!\!\!\!\!\!\!\!\!\!\!\!
\!\!\!\!\!\!\!\!\!\!\!\!\!\!\!\!\!\!\!\!\!C^{H^{0}}_{Q_{1}}
(\bar{\xi}^{U}_{N,tt}) 
& = & \frac{g^2 \bar{\xi}^{U}_{N,tt} \bar{\xi}^{D}_{N,bb} 
m_{\tau}}{64 \pi^2 m^2_{H^{0}} m_t } \Bigg{(}
\frac{(-1+2 \cos^2 \theta_W) \, y_t}{\cos^2 \theta_W}\Big{(}\frac{1}{\Theta_4
(y_t)}+\frac{z_H}{\Theta_7} \Big{)}-
\frac{x_t y_t}{\Theta_5 (y_t)}+\frac {x_t y_t(x y-z_H)}
{\Theta_1 (z_H,y_t)}  \nnb
\\ & & -2 x_t\, \ln \Bigg{[}
\frac{\Theta_5 (y_t) z_H} {\Theta_1 (z_H,y_t)} \Bigg{]} \Bigg{)}  ,
\nnb
\eea
\bea
\lefteqn{ C^{H^0}_{Q_{1}}(g^4) =-\frac{g^4 m_b m_{\tau} x_t}
{128 \pi^2 m^2_{H^{0}} m^2_t} 
\Bigg{(} -1+\frac{(-1+2x) x_t}{\Theta_5 (x_t) + y (1-x_t)}+
\frac{2 x_t (-1+ (2+x_t) y)}{\Theta_5 (x_t)} } \nnb \\
& & 
-\frac{4 \cos^2 \theta_W (-1+x+y)+ x_t(x+y)} {\cos^2 \theta_W 
\Theta_4 (x_t)} +\frac{x_t (x (x_t (y-2 z_H)-4 z_H)+2 z_H)} {\Theta_1 
(z_H,x_t)} \nnb \\ 
& &
+\frac{y_t ( (-1+x) x_t z_H+\cos^2 \theta_W ( (3 x-y) z_H+x_t 
(2 y (x-1)- z_H (2-3 x -y))))}{\cos^2 \theta_W (\Theta_3 (z_H)+x 
(x_t-y_t) z_H)} \nnb
\\ & & 
+ 2\, ( x_t \ln \Bigg{[} \frac{\Theta_5 (x_t) z_H}{\Theta_1 (z_H,x_t)} 
\Bigg{]}+ \ln \Bigg{[} \frac{x(y_t-x_t) z_H-\Theta_3 (z_H)} {(\Theta_5 (x_t)+ 
y (1-x_t) y_t z_H} \Bigg{]} )\Bigg{)}  ,\nnb  
\eea
\bea
C^{h_0}_{Q_1}((\bar{\xi}^U_{N,tt})^3) &=&
-\frac{\bar{\xi}^D_{N,\tau\tau} (\bar{\xi}^U_{N,tt})^3 m_b y_t}
{32 \pi^2 m_{h^O}^2 m_t \Theta_1 (z_h,y_t) \Theta_5 (y_t)}
 \Big{(} \Theta_1 (z_h,y_t) (2 y-1) + \Theta_5 (y_t) (2 x-1) z_h \Big{)}\, , \nnb 
\eea
\bea
C^{h_0}_{Q_1}((\bar{\xi}^U_{N,tt})^2) & = &
\frac{\bar{\xi}^D_{N,\tau\tau}\bar{\xi}^D_{N,bb} (\bar{\xi}^U_{N,tt})^2 }
{32 \pi^2  m_{h^O}^2  } \Bigg{(}
\frac{ (\Theta_5 (y_t) z_h (y_t-1)(x+y-1)-\Theta_1 (z_h,y_t)( \Theta_5(y_t)+y_t )
}{\Theta_1 (z_h)\Theta_5(y_t)}\nnb \\ & - &  2 \ln \Bigg{[} \frac{z_h \Theta_5
(y_t)}{\Theta_1 (z_h)} \Bigg{]} \Bigg{)} \, ,\nnb 
\eea
\bea
C^{h^0}_{Q_{1}}(\bar{\xi}^{U}_{N,tt}) & = & -\frac{g^2
\bar{\xi}^{D}_{N,\tau\tau}\bar{\xi}^{U}_{N,tt} m_b x_t}{64 \pi^2 m^2_{h^{0}} 
m_t} \Bigg{(}\frac{2 (-1+(2+x_t) y)}{\Theta_5 (x_t)}-\frac{x_t
(x-1)(y_t-z_h)}{\Theta'_2 (z_h)}+2 \ln \Bigg{[}\frac{z_h \Theta_5
(x_t)}{\Theta_1 (z_h,x_t)} \Bigg{]} \nnb \\ & + & \frac{x (x_t (y-2 z_h)-
4 z_h)+2 z_h}{\Theta_1 (z_h,x_t)}  -  \frac{(1+x) y_t z_h}{x y x_t y_t+z_h
((x-y)(x_t-y_t)- \Theta_6)} \nnb \\ 
& + & 
\frac{\Theta_9 + y_t z_h ( (x-y)(x_t-y_t)-\Theta_6 )(2
x-1)}{z_h \Theta_6 (\Theta_6 -(x-y)(x_t-y_t))}+\frac{x (y_t z_h + x_t
(z_h-y_t))-2 y_t z_h}{\Theta_2 (z_h)} \Bigg{)}, \nnb
\eea
\bea
C^{h^0}_{Q_{1}}(\bar{\xi}^{D}_{N,bb}) &  = & -\frac{g^2
\bar{\xi}^{D}_{N,\tau\tau}\bar{\xi}^{D}_{N,bb}}{64 \pi^2 m^2_{h^0} }
\Bigg{(}\frac{y x_t y_t (x x^2_t(y_t-z_h)+\Theta_6 z_h
(x-2))}{z_h\Theta_2 (z_h)\Theta_6 }+2 \ln
\Bigg{[}\frac{z_h\Theta_6}{\Theta_2 (z_h)} \Bigg{]}\Bigg{]}
\Bigg{)}\, , \nnb 
\eea
where 
\bea
\Theta_1 (\omega , \lambda ) & = & -(-1+y-y \lambda ) \omega -x (y \lambda
+\omega - \omega \lambda ) \nnb \\
\Theta_2 (\omega ) & = &  (x_t +y (1-x_t)) y_t \omega - x x_t (y
y_t+(y_t-1) \omega)   \nnb \\
\Theta^{\prime}_2  (\omega ) & = & \Theta_2 (\omega , x_t \leftrightarrow y_t)    \nnb \\
\Theta_3 (\omega) & = & (x_t (-1+y)-y ) y_t \omega +
x x_t (y y_t+\omega(-1+y_t)) \nnb \\
\Theta_4 (\omega) & = & 1-x +x  \omega  \nnb \\
\Theta_5 (\lambda) & = & x + \lambda (1-x) \nnb \\
\Theta_6  & = & (x_t +y  (1-x_t))y_t +x x_t  (1-y_t) \nnb \\
\Theta_7  & = & (y (y_t -1)-y_t) z_H+x (y y_t + (y_t-1) z_H ) \\ 
\Theta_8  & = & y_t (2 x^2 (1+x_t) (y_t-1) +x_t (y(1-y_t)+y_t)+x
(2(1-y+y_t) \nnb \\ & + & x_t (1-2 y (1-y_t)-3 y_t))) \nnb \\
\Theta_9  & = & -x^2_t (-1+x+y)(-y_t+x (2 y_t-1)) (y_t-z_h)-x_t y_t z_h
(x(1+2 x)-2 y) \nnb \\ & + & y^2_t (x_t (x^2 -y (1-x))+(1+x) (x-y) z_h) 
\nnb
\eea
and
\begin{eqnarray}
& & x_t=\frac{m_t^2}{m_W^2}~~~,~~~y_t=
\frac{m_t^2}{m_{H^{\pm}}}~~~,~~~z_H=\frac{m_t^2}{m^2_{H^0}}~~~,~~~
z_h=\frac{m_t^2}{m^2_{h^0}}~~~,~~~ z_A=\frac{m_t^2}{m^2_{A^0}}~~~,~~~ \nnb
\end{eqnarray}
The explicit forms of the functions $F_{1(2)}(y_t)$, $G_{1(2)}(y_t)$, 
$H_{1}(y_t)$ and $L_{1}(y_t)$ in eq.(\ref{CH}) are given as
\begin{eqnarray}
F_{1}(y_t)&=& \frac{y_t(7-5 y_t-8 y_t^2)}{72 (y_t-1)^3}+
\frac{y_t^2 (3 y_t-2)}{12(y_t-1)^4} \,\ln y_t \nonumber  \,\, , 
\\ 
F_{2}(y_t)&=& \frac{y_t(5 y_t-3)}{12 (y_t-1)^2}+
\frac{y_t(-3 y_t+2)}{6(y_t-1)^3}\, \ln y_t 
\nonumber  \,\, ,
\\ 
G_{1}(y_t)&=& \frac{y_t(-y_t^2+5 y_t+2)}{24 (y_t-1)^3}+
\frac{-y_t^2} {4(y_t-1)^4} \, \ln y_t
\nonumber  \,\, ,
\\ 
G_{2}(y_t)&=& \frac{y_t(y_t-3)}{4 (y_t-1)^2}+\frac{y_t} {2(y_t-1)^3} \, 
\ln y_t  \nonumber\,\, ,
\\
H_{1}(y_t)&=& \frac{1-4 sin^2\theta_W}{sin^2\theta_W}\,\, \frac{xy_t}{8}\,
\left[ \frac{1}{y_t-1}-\frac{1}{(y_t-1)^2} \ln y_t \right]\nonumber \\
&-&
y_t \left[\frac{47 y_t^2-79 y_t+38}{108 (y_t-1)^3}-
\frac{3 y_t^3-6 y_t+4}{18(y_t-1)^4} \ln y_t \right] 
\nonumber  \,\, , 
\\ 
L_{1}(y_t)&=& \frac{1}{sin^2\theta_W} \,\,\frac{x y_t}{8}\, 
\left[-\frac{1}{y_t-1}+ \frac{1}{(y_t-1)^2} \ln y_t \right]
\nonumber  \,\, .
\\ 
\label{F1G1}
\end{eqnarray}
Finally, the initial values of the coefficients in the model III are
\begin {eqnarray}   
C_i^{2HDM}(m_{W})&=&C_i^{SM}(m_{W})+C_i^{H}(m_{W}) , \nnb \\
C_{Q_{1}}^{2HDM}(m_{W})&=& \int^{1}_{0}dx \int^{1-x}_{0} dy \,
(C^{H^{0}}_{Q_{1}}((\bar{\xi}^{U}_{N,tt})^{2})+
 C^{H^{0}}_{Q_{1}}(\bar{\xi}^{U}_{N,tt})+
 C^{H^{0}}_{Q_{1}}(g^{4})+C^{h^{0}}_{Q_{1}}((\bar{\xi}^{U}_{N,tt})^{3}) \nnb
\\ & + &
 C^{h^{0}}_{Q_{1}}((\bar{\xi}^{U}_{N,tt})^{2})+
 C^{h^{0}}_{Q_{1}}(\bar{\xi}^{U}_{N,tt})+
 C^{h^{0}}_{Q_{1}}(\bar{\xi}^{D}_{N,bb})) , \nnb  \\
 C_{Q_{2}}^{2HDM}(m_{W})&=& \int^{1}_{0}dx \int^{1-x}_{0} dy\,
(C^{A^{0}}_{Q_{2}}((\bar{\xi}^{U}_{N,tt})^{3})+
C^{A^{0}}_{Q_{2}}((\bar{\xi}^{U}_{N,tt})^{2})+
 C^{A^{0}}_{Q_{2}}(\bar{\xi}^{U}_{N,tt})+
 C^{A^{0}}_{Q_{2}}(\bar{\xi}^{D}_{N,bb}))\nnb \\
C_{Q_{3}}^{2HDM}(m_W) & = & \frac{m_b}{m_{\tau} \sin^2 \theta_W} 
 (C_{Q_{1}}^{2HDM}(m_W)+C_{Q_{2}}^{2HDM}(m_W)) \nnb \\
C_{Q_{4}}^{2HDM}(m_W) & = & \frac{m_b}{m_{\tau} \sin^2 \theta_W} 
 (C_{Q_{1}}^{2HDM}(m_W)-C_{Q_{2}}^{2HDM}(m_W)) \nnb \\
C_{Q_{i}}^{2HDM} (m_W) & = & 0\,\, , \,\, i=5,..., 10.
\label{CiW}
\end{eqnarray}
Here, we present $C_{Q_{1}}$ and $C_{Q_{2}}$ in terms of the Feynmann
parameters $x$ and $y$ since the integrated results are extremely large.
Using these initial values, we can calculate the coefficients 
$C_{i}^{2HDM}(\mu)$ and $C^{2HDM}_{Q_i}(\mu)$ 
at any lower scale in the effective theory 
with five quarks, namely $u,c,d,s,b$ similar to the SM case 
\cite{buras,Chao,alil2,misiak}. 

The Wilson  coefficients playing  the essential role 
in this process are $C_{7}^{2HDM}(\mu)$, $C_{9}^{2HDM}(\mu)$,
$C_{10}^{2HDM}(\mu)$, 
$C^{2HDM}_{Q_1}(\mu )$ and $C^{2HDM}_{Q_2}(\mu )$. For completeness,
in the following we give their explicit expressions. 
\begin{eqnarray}
C_{7}^{eff}(\mu)&=&C_{7}^{2HDM}(\mu)+ Q_d \, 
(C_{5}^{2HDM}(\mu) + N_c \, C_{6}^{2HDM}(\mu)) \, \, , 
\label{C7eff}
\end{eqnarray}
where the LO  QCD corrected Wilson coefficient 
$C_{7}^{LO, 2HDM}(\mu)$ is given by
\begin{eqnarray} 
C_{7}^{LO, 2HDM}(\mu)&=& \eta^{16/23} C_{7}^{2HDM}(m_{W})+(8/3) 
(\eta^{14/23}-\eta^{16/23}) C_{8}^{2HDM}(m_{W})\nonumber \,\, \\
&+& C_{2}^{2HDM}(m_{W}) \sum_{i=1}^{8} h_{i} \eta^{a_{i}} \,\, , 
\label{LOwils}
\end{eqnarray}
and $\eta =\alpha_{s}(m_{W})/\alpha_{s}(\mu)$, $h_{i}$ and $a_{i}$ are 
the numbers which appear during the evaluation \cite{buras}. 

$C_9^{eff}(\mu)$ contains a perturbative part and a part coming from LD
effects due to conversion of the real $\bar{c}c$ into lepton pair $\tau^+
\tau^-$:
\begin{eqnarray}
C_9^{eff}(\mu)=C_9^{pert}(\mu)+ Y_{reson}(\hat{s})\,\, ,
\label{C9efftot}
\end{eqnarray}
where
\begin{eqnarray} 
C_9^{pert}(\mu)&=& C_9^{2HDM}(\mu) \nonumber 
\\ &+& h(z,  s) \left( 3 C_1(\mu) + C_2(\mu) + 3 C_3(\mu) + 
C_4(\mu) + 3 C_5(\mu) + C_6(\mu) \right) \nonumber \\
&- & \frac{1}{2} h(1, s) \left( 4 C_3(\mu) + 4 C_4(\mu) + 3
C_5(\mu) + C_6(\mu) \right) \\
&- &  \frac{1}{2} h(0,  s) \left( C_3(\mu) + 3 C_4(\mu) \right) +
\frac{2}{9} \left( 3 C_3(\mu) + C_4(\mu) + 3 C_5(\mu) + C_6(\mu)
\right) \nonumber \,\, ,
\label{C9eff2}
\end{eqnarray}
and
\begin{eqnarray}
Y_{reson}(\hat{s})&=&-\frac{3}{\alpha^2_{em}}\kappa \sum_{V_i=\psi_i}
\frac{\pi \Gamma(V_i\rightarrow \tau^+ \tau^-)m_{V_i}}{q^2-m_{V_i}+i m_{V_i}
\Gamma_{V_i}} \nonumber \\
& & \left( 3 C_1(\mu) + C_2(\mu) + 3 C_3(\mu) + 
C_4(\mu) + 3 C_5(\mu) + C_6(\mu) \right).
\label{Yres}
\end{eqnarray}
In eq.(\ref{C9efftot}), the functions $h(u, s)$ are given by
\begin{eqnarray}
h(u, s) &=& -\frac{8}{9}\ln\frac{m_b}{\mu} - \frac{8}{9}\ln u +
\frac{8}{27} + \frac{4}{9} x \\
& & - \frac{2}{9} (2+x) |1-x|^{1/2} \left\{\begin{array}{ll}
\left( \ln\left| \frac{\sqrt{1-x} + 1}{\sqrt{1-x} - 1}\right| - 
i\pi \right), &\mbox{for } x \equiv \frac{4u^2}{ s} < 1 \nonumber \\
2 \arctan \frac{1}{\sqrt{x-1}}, & \mbox{for } x \equiv \frac
{4u^2}{ s} > 1,
\end{array}
\right. \\
h(0,s) &=& \frac{8}{27} -\frac{8}{9} \ln\frac{m_b}{\mu} - 
\frac{4}{9} \ln s + \frac{4}{9} i\pi \,\, , 
\label{hfunc}
\end{eqnarray}
with $u=\frac{m_c}{m_b}$.
The phenomenological parameter $\kappa$ in eq. (\ref{Yres}) is taken as 
$2.3$. In eqs. (30) and (\ref{Yres}), the contributions of 
the coefficients $C_1(\mu)$, ...., $C_6(\mu)$ are due to the operator mixing.

Finally, the Wilson coefficients $C_{Q_1}(\mu)$ and $C_{Q_2}(\mu )$  
are given by \cite{Chao}
\beq
C_{Q_i}(\mu )=\eta^{-12/23}\,C_{Q_i}(m_W)~,~i=1,2~. 
\eeq
\end{appendix}
\newpage

\end{document}